\documentclass{aa}
\input {epsf}
\usepackage{graphicx}
\input psfig.sty

\begin{document}

\title{The abundance pattern of O, Mg, Si and Fe in the intracluster medium of
the Centaurus cluster observed with  XMM-Newton}
\titlerunning{The abundance pattern of the ICM of the Centaurus cluster}
\offprints{K. Matsushita\\ \email{matusita@rs.kagu.tus.ac.jp}}
\date{}
\author{Kyoko Matsushita\inst{1,2}, Hans B\"ohringer\inst{2}, Isao
Takahashi\inst{3},Yasushi Ikebe\inst{4}}
\authorrunning{Kyoko Matsushita et al.}
\institute{
Department of Physics, Tokyo University of Science, 1-3 Kagurazaka,
Shinjyuku-ku, Tokyo, 162-8601 Japan 
 \and Max-Planck-Institut f\"ur Extraterrestrial Physik, D-85748 Garching, 
Germany
\and 
Department of Physics, University of Tokyo, 7-3-1 Hongo, Bunkyo-ku,
Tokyo, 113-8654
Japan, \and
Joint Center for Astrophysics, University of Maryland, Baltimore County,
1000 Hilltop Circle, Baltimore, MD 21250, 
USA}
\abstract{
The abundances of O, Mg, Si and Fe in the intracluster medium of the
Centaurus cluster are derived.
 The Fe abundance has a negative radial gradient.
In  solar units, the Si abundance is close to the Fe abundance, while 
  the O and Mg abundances are much smaller.
The high Fe/O and  Si/O ratio indicate that metal supply
from  supernovae Ia  is
important  and  supernovae Ia synthesize Si as well as Fe.
 Within 2$'$, the O and Mg abundances are consistent with the stellar metallicity
of the cD galaxy derived from the Mg$_2$ index.
This result indicates that the central gas is dominated by the gas from
the cD galaxy.
The observed abundance pattern of the Centaurus cluster resembles to
 those observed in center of other clusters and groups of galaxies.
However, 
the central Fe abundance and the Si/Fe ratio are 40 \% higher and 30\% smaller than
those of M 87, respectively. 
Since the accumulation timescale of the supernovae Ia
is higher in the Centaurus cluster, these differences imply a time
 dependence of  nucleosynthesis by supernovae Ia.
\keywords{X-rays:ICM --- galaxies:ISM --- individual:Centaurus cluster}
}
\maketitle

\section{Introduction}

The intracluster medium (ICM) contains a large amount of metals, which
are synthesized by supernovae (SNe) Ia and SNe II in cluster galaxies.
A knowledge of the contributions from the two types of SNe to the metal abundances are important
to  understand
the origin of the metals and to study the evolution of galaxies in clusters.

The XMM-Newton observatory  enables us to derive O and Mg abundances, which
are not synthesized by SN Ia.
For the ICM of M 87 observed with the  XMM,
the O/Si ratio is found to be about a half of the solar ratio, while 
Fe and Si have  similar values in units of the solar abundance
(B\"ohringer et al. 2001; Finoguenov et al. 2002; Matsushita et al. 2003).
Such low O abundances compared to the Si abundances are also
derived for the centers of other clusters and groups of galaxies
(e.g. Tamura et al. 2001b ; Boute et al. 2003, and others).

In order to  account for the observed O/Si/Fe abundance pattern of M 87,
the Fe/Si ratio of ejecta of SN Ia should be about 1
solar ratio. This value is significantly higher than expected 
and probably  related with dimmer SNe Ia observed in old
stellar systems (Hamuy et al. 1996; Ivanov et al. 2000).
Since the mass of synthesized Ni$^{56}$ determines the
luminosity of each SN, 
the abundance ratio in the ejecta may be correlated to  the luminosity
of SN Ia.
 In this context a lower Ni$^{56}$ yield implies less complete
nuclear burning with a higher ratio of $\alpha$-elements such as Si to Fe.
The observed O/Si/Fe pattern of the Galactic stars
 also indicates that the Si/Fe ratio in ejecta of SN Ia depends on stellar 
metallicity, i.e., age of  the system (Matsushita et al. 2003)

The variation in the observed abundance pattern reflects variations in the
 explosion models, e.g. a high fraction of Fe for the classical deflagration model, W7
(Nomoto et al. 1984) or a larger fraction of Si for the delayed detonation model, WDDs
(Nomoto et al. 1997; Iwamoto et al. 1999).
The W7 model,  which can well reproduce observed optical spectra of SNe Ia 
(Nugent et al. 1997), predicts a high Fe/Si ratio of 2.6 in units of the
solar ratio.
In contrast, the WDD models give a wider range of 
Fe/Si ratios from 1 to 3 in units of the solar ratio, which might be related to the age of
the system (Umeda et al. 1999).

In this paper, we report on the O/Mg/Si/Fe abundance pattern of the Centaurus
cluster observed with XMM.
 The temperature properties of the cluster are described in Takahashi (2004).
The Centaurus cluster is a nearby cluster, centered on the cD galaxy, 
 NGC 4696.
A positive temperature gradient and a negative abundance gradient are observed
with ASCA  (Fukazawa et al.1994; Ikebe et al.1999).
The cluster is expected to have a cooling flow  (Allen \& Fabian 1994),
and Chandra observed an  unusual plume-like cool structure at the center
(Sanders \& Fabian 2002).
 However,  as in other cooling flow clusters,
 the luminosity of the central cool component is much
smaller than the expected value from the cooling flow model (e.g. Makishima
et al. 2001; Tamura et al. 2001a; B\"ohringer et al. 2002; Matsushita et
al. 2002).
The problem of the missing of the cooling component becomes an advantage in
determination of elemental abundances, since the observed narrow temperature range of the ICM
 strongly reduces the ambiguity in the abundance measurement.

We adopt for the solar abundances the values given by Feldman
(1992), where the solar  Fe abundance relative to H is
3.24$\times10^{-5}$  in number.
 A value for the Hubble constant of $H_0=70 ~\rm{km/s/Mpc}$ is assumed.
Unless otherwise specified, we use 90\% confidence error regions.

\section{Observation}

The Centaurus cluster was observed with XMM-Newton on January 3rd, 2002.
 The thin filter is used for the EMOS and the EPN.
We have selected events with pattern smaller than 5 and 13 for the EPN and
the EMOS, respectively.
The background spectrum was calculated for each spectrum
by integrating blank sky data in the same detector regions.
Among deep sky observations collected with  XMM, we have selected the data
with the most similar background to that of the Centaurus cluster,
after screening  background flare events in the data and the background
following Katayama et al. (2004).
The effective exposures of the EPN and the EMOS are 29 ks and 32 ks, respectively.
The spectra  are  accumulated within rings, centered on the X-ray peak of the cluster.
After subtracting the background and correcting for the effect of  vignetting,
 deprojected spectra are calculated  assuming spherical symmetry.
For the EPN spectra, the response matrix  file  and the auxiliary response file  corresponding
to each spectrum are calculated using  SAS-v5.4.1.
For the EMOS data,  we used m1\_thin1v9q20t5r6\_all\_15.rsp.
The spectral analysis uses the XSPEC\_v11.2.0 package.
For the EPN data, we used the spectral range from 0.5 to 7.3 keV,
in order to discard strong instrumental lines around 8 keV.
We fitted  the  EMOS data in the spectral range from 0.4 to 9.0 keV.
 At $r>8'$, the energy range of 1.45--1.55 keV of the EMOS 
is discarded
due to a  strong instrumental line.
Here, $r$ is the projected radius from the X-ray peak.
We also denote $R$ as the deprojected or three-dimensional radius.

\section{Spectral fit with the MEKAL model}

 The first step of the analysis is fitting the annular (projected) spectra
with a single temperature MEKAL  (e.g. Kaastra  1992; Liedahl et
al. 1995) model, 
and the  deprojected spectra  with a single and two-temperature MEKAL
model with photoelectric absorption.
Abundances of C and N are fixed to be 1 solar and those of other
elements are determined separately.
The abundances of each element for all temperature
components are assumed to have the same values.
In addition, the deprojected spectra are fitted with a multi-temperature model which
is a sum of 14 temperature components.
 Since a spectrum from  an isothermal plasma can be well reproduced by a sum of
spectra of plasma of two neighboring temperatures  with similar elemental abundances,
we selected temperatures in order to derive  metal abundances with an
accuracy within 5 \%.
Thus, in order to determine the abundance distribution, this multi-temperature 
model is enough to fit any 
temperature distribution at least when all temperature components have
the same abundances.

\begin{figure}[]
\resizebox{8.1cm}{!}{\includegraphics{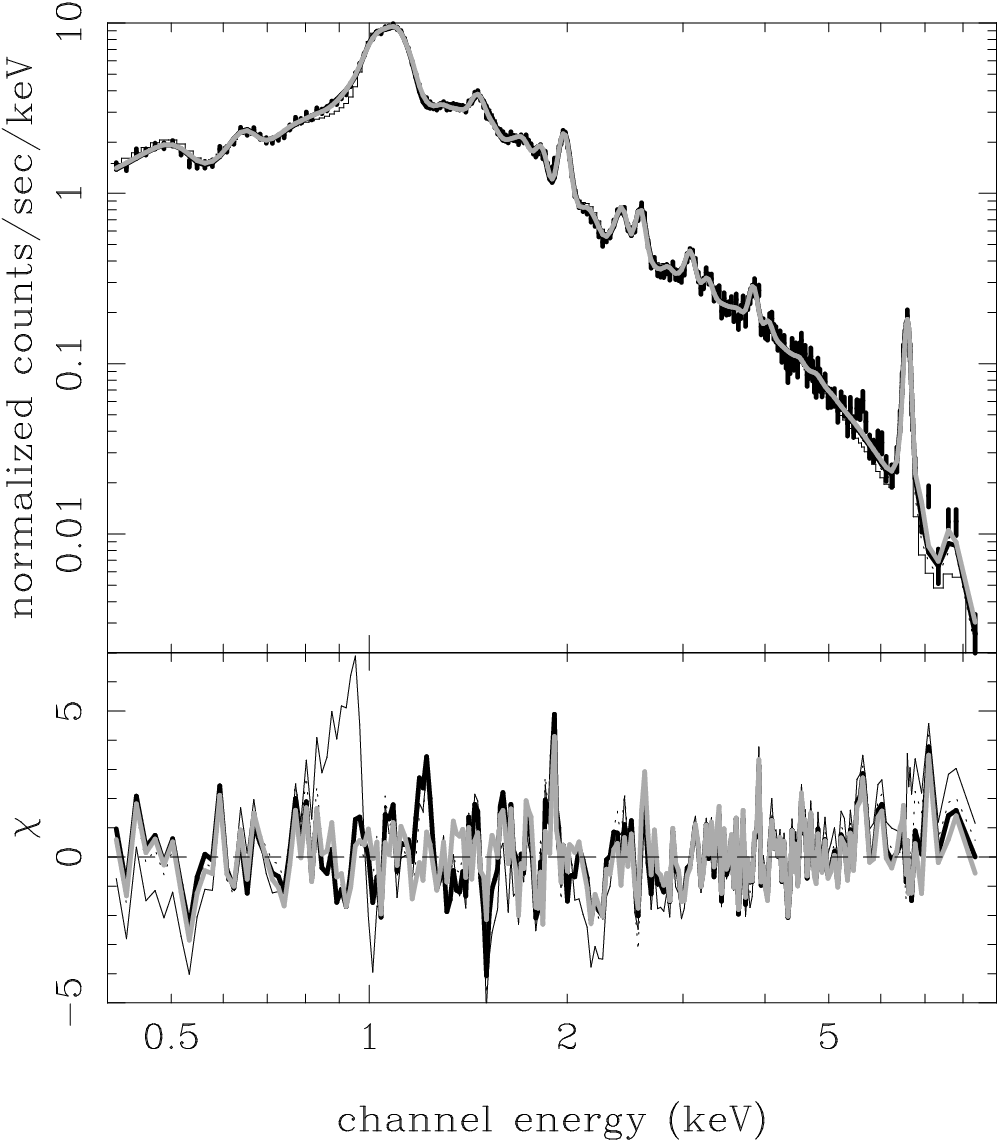}}
\caption{The deprojected spectrum at $R$=0.5--2$'$ of the EMOS1+EMOS2 of the Centaurus cluster fitted with
 the single temperature MEKAL model (black thin lines), the
 two-temperature MEKAL model (black dotted lines), 
 the multi-temperature MEKAL model (black bold lines), and the 
 multi-temperature APEC model (gray bold lines).
}
\label{depro_abund}
\end{figure}

\begin{table*}
\caption{Results of spectrum fitting of the projected spectra.}
 \begin{tabular}[t]{rlllllllllr}
 $r$     &  &   & kT &   & $N_{\rm H}$ & O & Mg & Si  & Fe &  $\chi^2/dof$\\
(arcmin)& && (keV) & (keV) & (10$^{20}{\rm cm^{-2}}$) & (solar)&
  (solar)& (solar)& (solar)& \\\hline
  0.05- 0.25 &EMOS &1T &1.00& &17.8& 0.00& 0.09& 0.28& 0.34& 837/ 61 \\
 0.25- 0.50 &EMOS &1T &1.36& &15.0& 0.25& 0.00& 0.47& 0.67& 740/ 96 \\
 0.50- 0.70 &EMOS &1T &1.66& &11.8& 0.33& 0.09& 1.01& 1.20& 386/ 93 \\
 0.70- 1.00 &EMOS &1T &1.89& &10.9& 0.44& 0.18& 1.31& 1.50& 483/127 \\
 1.00- 1.40 &EMOS &1T &2.42$^{+0.05}_{-0.05}$ &&9.4$^{+0.6}_{-0.6}$& 0.64$^{+0.19}_{-0.18}$& 0.28& 1.60$^{+0.17}_{-0.16}$& 2.04$^{+0.13}_{-0.12}$& 283/148 \\
 1.40- 2.00 &EMOS &1T &2.77$^{+0.05}_{-0.05}$ &&8.8$^{+0.5}_{-0.5}$& 0.50$^{+0.17}_{-0.16}$& 0.41& 1.58$^{+0.15}_{-0.14}$& 1.83$^{+0.09}_{-0.09}$& 266/169 \\
 2.00- 2.80 &EMOS &1T &3.00$^{+0.05}_{-0.05}$ &&8.8$^{+0.4}_{-0.5}$& 0.70$^{+0.17}_{-0.16}$& 0.56& 1.43$^{+0.14}_{-0.13}$& 1.66$^{+0.08}_{-0.08}$& 307/193 \\
 2.80- 4.00 &EMOS &1T &3.32$^{+0.06}_{-0.06}$ &&9.2$^{+0.4}_{-0.4}$& 0.60$^{+0.15}_{-0.15}$& 0.31& 0.98$^{+0.12}_{-0.12}$& 1.14$^{+0.06}_{-0.06}$& 331/224 \\
 4.00- 5.60 &EMOS &1T &3.60$^{+0.07}_{-0.07}$ &&8.2$^{+0.4}_{-0.3}$& 0.37$^{+0.14}_{-0.14}$& 0.48& 0.78$^{+0.11}_{-0.11}$& 0.80$^{+0.05}_{-0.05}$& 329/240 \\
 5.60- 8.00 &EMOS &1T &3.71$^{+0.10}_{-0.08}$ &&8.0$^{+0.4}_{-0.4}$& 0.49$^{+0.15}_{-0.15}$& 0.48& 0.70$^{+0.11}_{-0.11}$& 0.68$^{+0.05}_{-0.05}$& 366/271 \\
 8.00-13.50 &EMOS &1T &3.75$^{+0.12}_{-0.10}$ &&7.3$^{+0.4}_{-0.4}$& 0.43$^{+0.14}_{-0.13}$& 0.84& 0.62$^{+0.11}_{-0.10}$& 0.57$^{+0.05}_{-0.05}$& 439/325 \\
  8.00-13.50 &EMOS &2T &4.30$^{+1.68}_{-0.29}$ &1.43$^{+0.76}_{-0.21}$  &7.3$^{+0.3}_{-0.4}$& 0.44$^{+0.07}_{-0.13}$& 0.73& 0.65$^{+0.12}_{-0.13}$& 0.53$^{+0.08}_{-0.11}$& 409/322 \\
  8.00-13.50 &EMOS &14T& & &9.3$^{+1.2}_{-1.3}$& 0.33$^{+0.16}_{-0.08}$& 0.78& 0.66$^{+0.09}_{-0.09}$& 0.57$^{+0.04}_{-0.06}$& 396/311 \\
\hline
 0.05- 0.25 &EPN &1T &0.78& &28.8& 0.01& 0.00& 0.09& 0.14& 1151/139 \\
 0.25- 0.50 &EPN &1T &1.37& &12.5& 0.00& 0.00& 0.45& 0.71& 1087/201 \\
 0.50- 1.00 &EPN &1T &1.89& &8.7& 0.04& 0.00& 0.98& 1.48& 1044/343 \\
 1.00- 1.40 &EPN &1T &2.38$^{+0.04}_{-0.04}$ &&7.2$^{+0.6}_{-0.7}$& 0.19$^{+0.15}_{-0.13}$& 0.00& 1.38$^{+0.17}_{-0.13}$& 1.88$^{+0.11}_{-0.08}$& 476/313 \\
 1.40- 2.00 &EPN &1T &2.70$^{+0.04}_{-0.04}$ &&7.2$^{+0.5}_{-0.5}$& 0.25$^{+0.13}_{-0.13}$& 0.08& 1.29$^{+0.14}_{-0.13}$& 1.68$^{+0.08}_{-0.07}$& 459/364 \\
 2.00- 2.80 &EPN &1T &2.92$^{+0.05}_{-0.04}$ &&7.5$^{+0.5}_{-0.6}$& 0.39$^{+0.15}_{-0.13}$& 0.09& 1.00$^{+0.12}_{-0.11}$& 1.49$^{+0.08}_{-0.06}$& 579/412 \\
 2.80- 4.00 &EPN &1T &3.25$^{+0.05}_{-0.05}$ &&7.3$^{+0.5}_{-0.4}$& 0.06$^{+0.11}_{-0.06}$& 0.20& 0.82$^{+0.11}_{-0.11}$& 1.06$^{+0.05}_{-0.05}$& 533/472 \\
 4.00- 5.60 &EPN &1T &3.60$^{+0.06}_{-0.04}$ &&6.6$^{+0.4}_{-0.4}$& 0.12$^{+0.12}_{-0.12}$& 0.01& 0.54$^{+0.12}_{-0.12}$& 0.80$^{+0.04}_{-0.04}$& 591/496 \\
 5.60- 8.00 &EPN &1T &3.84$^{+0.07}_{-0.07}$ &&6.1$^{+0.4}_{-0.4}$& 0.05$^{+0.12}_{-0.05}$& 0.05& 0.40$^{+0.11}_{-0.11}$& 0.65$^{+0.04}_{-0.04}$& 579/527 \\
 8.00-16.00 &EPN &1T &3.98$^{+0.14}_{-0.05}$ &&5.7$^{+0.3}_{-0.5}$& 0.04$^{+0.16}_{-0.04}$& 0.00& 0.36$^{+0.11}_{-0.09}$& 0.63$^{+0.06}_{-0.03}$& 805/687 \\
 8.00-16.00 &EPN &2T &4.14$^{+0.17}_{-0.06}$ &1.06$^{+0.08}_{-0.17}$  &6.2$^{+0.4}_{-0.5}$& 0.24$^{+0.14}_{-0.12}$& 0.01& 0.44$^{+0.11}_{-0.11}$& 0.66$^{+0.05}_{-0.03}$& 746/684 \\
 8.00-16.00 &EPN &14T& & &6.4$^{+3.0}_{-0.4}$& 0.25$^{+0.18}_{-0.13}$& 0.00& 0.45$^{+0.09}_{-0.10}$& 0.66$^{+0.04}_{-0.05}$& 736/674 \\
\hline
\end{tabular}
\end{table*}
\begin{table*}
\caption{Results of spectrum fitting of the deprojected spectra}
 \begin{tabular}[t]{llllllllllr}
 $R$     & &   & kT1    & kT2  & $N_{\rm H}$ & O & Mg & Si  & Fe &  $\chi^2/dof$\\
(arcmin) & &&(keV) & (keV)  &\hspace{-0.3cm} (10$^{20}{\rm \small cm^{-2}}$) & (solar)&  (solar)& (solar)& (solar)& \\\hline
 0.00- 0.50 &EMOS &1T &1.08& &15.9& 0.17& 0.11& 0.35& 0.45& 1111/129 \\
0.00- 0.50 &EMOS &2T & 1.51$^{+0.05}_{-0.07}$ & 0.71$^{+0.01}_{-0.02}$   &11.4$^{+2.4}_{-1.1}$& 0.49$^{+0.18}_{-0.18}$& 0.89& 1.30$^{+0.23}_{-0.30}$& 1.57$^{+0.24}_{-0.28}$& 169/126 \\
 0.00- 0.50 &EMOS &14T& & &11.4$^{+3.7}_{-2.1}$& 0.48$^{+0.25}_{-0.19}$& 0.89& 1.39$^{+0.46}_{-0.33}$& 1.79$^{+0.42}_{-0.34}$& 139/116 \\
 0.00- 0.50 &EPN &1T &0.97& &16.9& 0.14& 0.04& 0.28& 0.38& 1167/176 \\
 0.00- 0.50 &EPN &2T &1.45$^{+0.07}_{-0.08}$ &0.70$^{+0.02}_{-0.02}$  &11.6$^{+1.2}_{-2.4}$& 0.59$^{+0.29}_{-0.19}$& 0.80& 1.52$^{+0.45}_{-0.38}$& 1.85$^{+0.29}_{-0.35}$& 195/173 \\
 0.00- 0.50 &EPN &14T& & &20.8$^{+12.9}_{-7.5}$& 0.71$^{+0.37}_{-0.28}$& 1.39& 2.48$^{+0.74}_{-1.00}$& 3.44$^{+0.74}_{-1.26}$& 170/163 \\
\hline
 0.50- 1.00 &EMOS &1T &1.43& &13.5& 0.49& 0.24& 1.20& 1.17& 354/146 \\
 0.50- 1.00 &EMOS &2T &1.74$^{+0.07}_{-0.03}$ &0.85$^{+0.03}_{-0.04}$  &9.7$^{+1.3}_{-1.6}$& 1.15$^{+0.45}_{-0.30}$& 1.26& 2.79$^{+0.72}_{-0.42}$& 2.71$^{+0.57}_{-0.33}$& 189/143 \\
 0.50- 1.00 &EMOS &14T& & &9.3$^{+1.2}_{-1.6}$& 1.22$^{+0.47}_{-0.33}$& 1.43& 3.06$^{+0.72}_{-0.62}$& 2.93$^{+0.56}_{-0.75}$& 189/133 \\
\hline
 1.00- 2.00 &EMOS &1T &2.30$^{+0.06}_{-0.05}$ &&8.8$^{+0.8}_{-0.8}$& 0.49$^{+0.22}_{-0.20}$& 0.34& 1.83$^{+0.22}_{-0.20}$& 2.22$^{+0.18}_{-0.16}$& 246/212 \\
 1.00- 2.00 &EMOS &2T &2.46$^{+0.25}_{-0.12}$ &1.30$^{+0.31}_{-0.37}$  &8.7$^{+0.8}_{-0.8}$& 0.62$^{+0.24}_{-0.22}$& 0.53& 2.07$^{+0.27}_{-0.25}$& 2.42$^{+0.23}_{-0.20}$& 220/209 \\
 1.00- 2.00 &EMOS &14T& & &8.6$^{+1.0}_{-0.8}$& 0.62$^{+0.25}_{-0.21}$& 0.58& 2.12$^{+0.28}_{-0.26}$& 2.44$^{+0.22}_{-0.21}$& 214/199 \\
\hline
 0.50- 2.00 &EMOS &1T &1.95& &10.2& 0.40& 0.20& 1.42& 1.70& 658/249 \\
 0.50- 2.00 &EMOS &2T &2.20$^{+0.07}_{-0.05}$ &1.00$^{+0.12}_{-0.08}$  &9.0$^{+0.6}_{-0.6}$& 0.78$^{+0.16}_{-0.16}$& 0.79& 2.23$^{+0.20}_{-0.20}$& 2.46$^{+0.16}_{-0.16}$& 339/246 \\
 0.50- 2.00 &EMOS &14T& & &8.9$^{+0.6}_{-0.7}$& 0.78$^{+0.17}_{-0.15}$& 0.80& 2.33$^{+0.21}_{-0.21}$& 2.54$^{+0.16}_{-0.17}$& 322/236 \\
 0.50- 2.00 &EPN &1T &2.10& &7.6& 0.08& 0.00& 1.24& 1.73& 1074/458 \\
 0.50- 2.00 &EPN &2T &2.21$^{+0.03}_{-0.03}$ &0.92$^{+0.05}_{-0.07}$  &7.9$^{+0.5}_{-0.6}$& 0.45$^{+0.13}_{-0.09}$& 0.19& 1.70$^{+0.16}_{-0.10}$& 2.24$^{+0.14}_{-0.07}$& 649/455 \\
 0.50- 2.00 &EPN &14T& & &8.0$^{+0.4}_{-0.5}$& 0.46$^{+0.11}_{-0.10}$& 0.17& 1.68$^{+0.13}_{-0.13}$& 2.14$^{+0.12}_{-0.09}$& 634/445 \\
\hline
 2.00- 4.00 &EMOS &1T &2.94$^{+0.06}_{-0.06}$ &&9.5$^{+0.6}_{-0.5}$& 0.78$^{+0.20}_{-0.19}$& 0.33& 1.42$^{+0.16}_{-0.15}$& 1.72$^{+0.10}_{-0.09}$& 342/294 \\
 2.00- 4.00 &EMOS &2T  &  5.30$^{+4.35}_{-2.14}$ & 2.42$^{+0.31}_{-0.89}$ &9.4$^{+0.6}_{-0.5}$& 0.72$^{+0.19}_{-0.18}$& 0.39& 1.39$^{+0.16}_{-0.15}$& 1.63$^{+0.13}_{-0.11}$& 324/291 \\
 2.00- 4.00 &EMOS &14T& & &9.3$^{+0.9}_{-0.4}$& 0.78$^{+0.20}_{-0.20}$& 0.44& 1.45$^{+0.15}_{-0.18}$& 1.67$^{+0.09}_{-0.12}$& 319/281 \\
 2.00- 4.00 &EPN &1T &2.84$^{+0.05}_{-0.05}$ &&7.7$^{+0.7}_{-0.7}$& 0.26$^{+0.17}_{-0.16}$& 0.22& 1.15$^{+0.17}_{-0.15}$& 1.56$^{+0.09}_{-0.08}$& 491/463 \\
 2.00- 4.00 &EPN &2T & 37$(>5.0)$  &  2.74$^{+0.08}_{-0.07}$ &7.8$^{+0.6}_{-0.8}$& 0.25$^{+0.16}_{-0.15}$& 0.18& 1.13$^{+0.19}_{-0.13}$& 1.54$^{+0.11}_{-0.07}$& 472/460 \\
 2.00- 4.00 &EPN &14T& & &8.3$^{+1.7}_{-0.9}$& 0.34$^{+0.17}_{-0.19}$& 0.19& 1.11$^{+0.18}_{-0.14}$& 1.48$^{+0.10}_{-0.08}$& 479/450 \\

\hline
 4.00- 8.00 &EMOS &1T &3.54$^{+0.11}_{-0.09}$ &&8.9$^{+0.6}_{-0.5}$& 0.43$^{+0.20}_{-0.19}$& 0.34& 0.78$^{+0.15}_{-0.14}$& 0.83$^{+0.07}_{-0.06}$& 391/332 \\
 4.00- 8.00 &EMOS &2T &6.32$^{+7.99}_{-4.15}$ &2.76$^{+0.55}_{-0.58}$  &8.8$^{+0.6}_{-0.6}$& 0.41$^{+0.19}_{-0.18}$& 0.32& 0.76$^{+0.15}_{-0.14}$& 0.80$^{+0.10}_{-0.09}$& 377/329 \\
 4.00- 8.00 &EMOS &14T& & &9.1$^{+1.0}_{-0.7}$& 0.41$^{+0.18}_{-0.18}$& 0.35& 0.78$^{+0.16}_{-0.14}$& 0.80$^{+0.09}_{-0.07}$& 377/319 \\
 4.00- 8.00 &EPN &1T &3.56$^{+0.10}_{-0.10}$ &&6.4$^{+0.8}_{-0.6}$& 0.03$^{+0.20}_{-0.03}$& 0.33& 0.58$^{+0.18}_{-0.17}$& 0.82$^{+0.07}_{-0.06}$& 541/534 \\
 4.00- 8.00 &EPN &2T &3.57$^{+0.12}_{-0.09}$ &0.84$^{+0.53}_{-0.36}$  &6.9$^{+0.8}_{-0.9}$& 0.12$^{+0.27}_{-0.12}$& 0.36& 0.60$^{+0.21}_{-0.16}$& 0.83$^{+0.11}_{-0.06}$& 536/531 \\
 4.00- 8.00 &EPN &14T& & &6.9$^{+1.5}_{-0.9}$& 0.15$^{+0.22}_{-0.15}$& 0.38& 0.62$^{+0.17}_{-0.17}$& 0.82$^{+0.07}_{-0.07}$& 535/521 \\
\hline
 2.00- 8.00 &EMOS &1T &3.26$^{+0.06}_{-0.05}$ &&9.3$^{+0.3}_{-0.3}$& 0.55$^{+0.12}_{-0.12}$& 0.33& 0.99$^{+0.09}_{-0.09}$& 1.14$^{+0.05}_{-0.05}$& 539/387 \\
 2.00- 8.00 &EMOS &2T &6.17$^{+2.56}_{-1.11}$ &2.60$^{+0.28}_{-0.24}$  &9.2$^{+0.3}_{-0.3}$& 0.52$^{+0.11}_{-0.11}$& 0.32& 0.97$^{+0.08}_{-0.08}$& 1.10$^{+0.07}_{-0.06}$& 495/384 \\
 2.00- 8.00 &EMOS &14T& & &9.2$^{+0.6}_{-0.3}$& 0.52$^{+0.13}_{-0.11}$& 0.34& 0.98$^{+0.08}_{-0.09}$& 1.08$^{+0.05}_{-0.05}$& 502/374 \\
 2.00- 8.00 &EPN &1T &3.23$^{+0.08}_{-0.02}$ &&7.1$^{+0.3}_{-0.7}$& 0.13$^{+0.13}_{-0.13}$& 0.28& 0.83$^{+0.17}_{-0.10}$& 1.15$^{+0.06}_{-0.03}$& 825/686 \\
 2.00- 8.00 &EPN &2T &3.18$^{+0.05}_{-0.04}$ &0.86$^{+0.37}_{-0.22}$  &7.6$^{+0.4}_{-0.7}$& 0.19$^{+0.17}_{-0.10}$& 0.27& 0.82$^{+0.15}_{-0.08}$& 1.12$^{+0.07}_{-0.03}$& 809/684 \\
 2.00- 8.00 &EPN &14T& & &7.7$^{+0.7}_{-1.0}$& 0.22$^{+0.13}_{-0.14}$& 0.26& 0.80$^{+0.06}_{-0.09}$& 1.05$^{+0.07}_{-0.04}$& 801/674 \\
\hline

\end{tabular}

\end{table*}

Table 1 and 2 summarize the results.
Even for the deprojected spectra, the single temperature model 
is not acceptable at  $R<2'$. 
The two-temperature model dramatically reduces  the $\chi^2$, and the multi-temperature
model slightly improves $\chi^2$.
Outside this radius, $\chi^2$ derived from the two- and the multi-temperature
model are slightly smaller than those of the single-temperature model.
Figure 1 shows a representative spectrum for $R$=0.5--2$'$.
Using the two- and the multi-temperature MEKAL model, not only the continuum and 
the Fe-L bump, but also the strength of  the Fe-K line is fitted well.
The consistency between the Fe-L and the Fe-K will be discussed in
detail at \S\ref{FeK}.
There remains a residual structure around 1.2--1.5 keV at the Fe-L/Mg-K region.
Therefore, in Table 1 and 2,  we do not show  errors in the  Mg abundances.

\begin{figure*}[]
\resizebox{8.1cm}{!}{\includegraphics{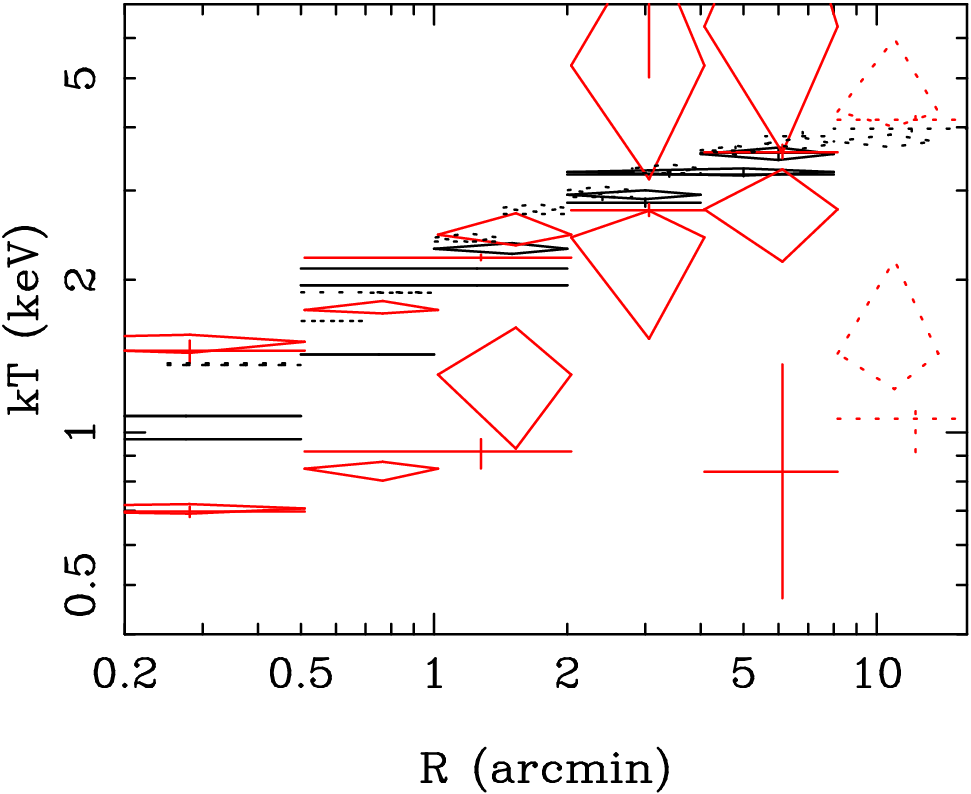}}
\resizebox{8.1cm}{!}{\includegraphics{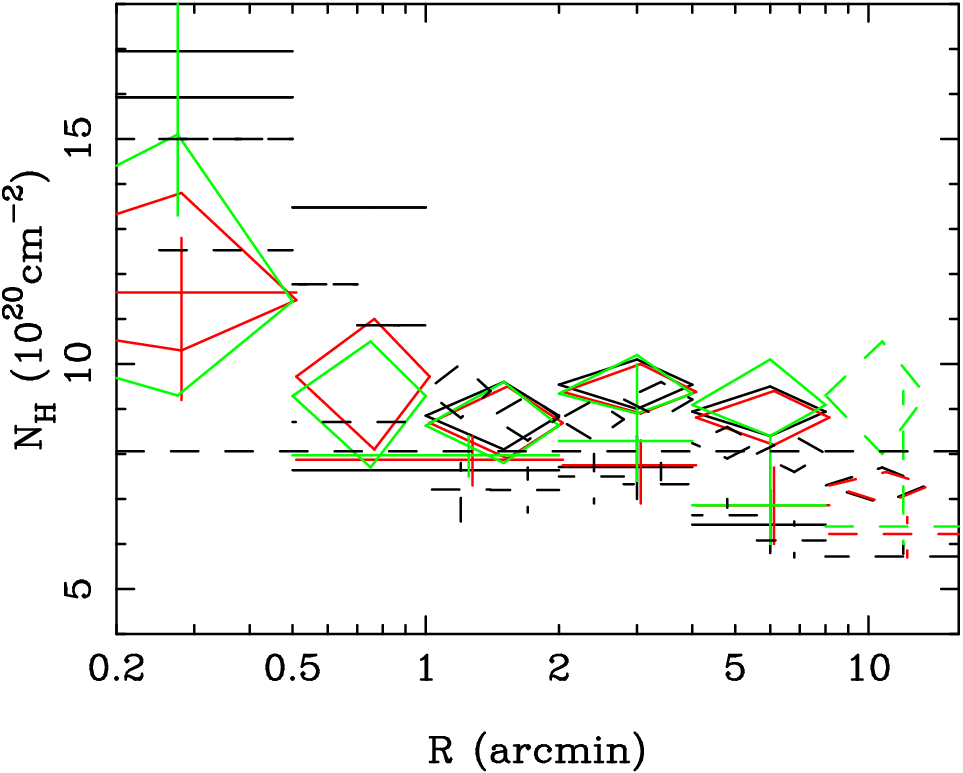}}
\caption{The radial profiles of the temperature and the hydrogen column density. 
Dotted and solid lines correspond to the results of the projected
 spectra and the deprojected spectra, respectively.
Black, red and green correspond to the single, two, and the
 multi-temperature MEKAL model, respectively.
The results from the EMOS are shown in diamonds and those from the EPN are
 shown in crosses.
The dashed horizontal line in the profile of the hydrogen column density
 represents the Galactic value.}
\label{ktnh}
\end{figure*}
\begin{figure}
\resizebox{8.1cm}{!}{\includegraphics{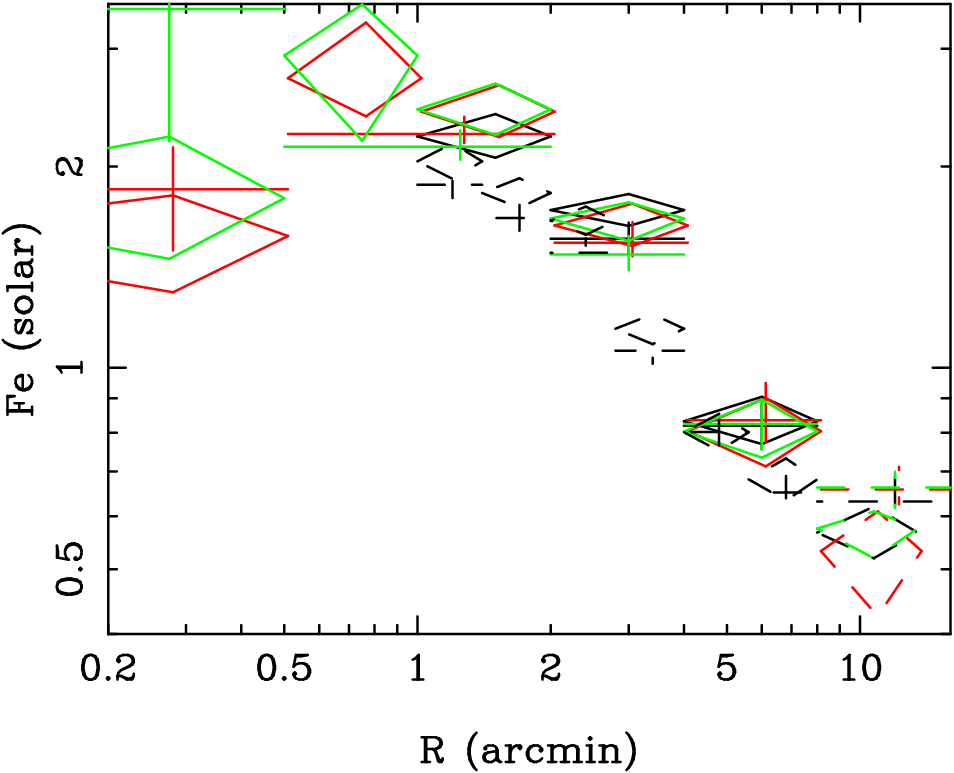}}
\resizebox{8.1cm}{!}{\includegraphics{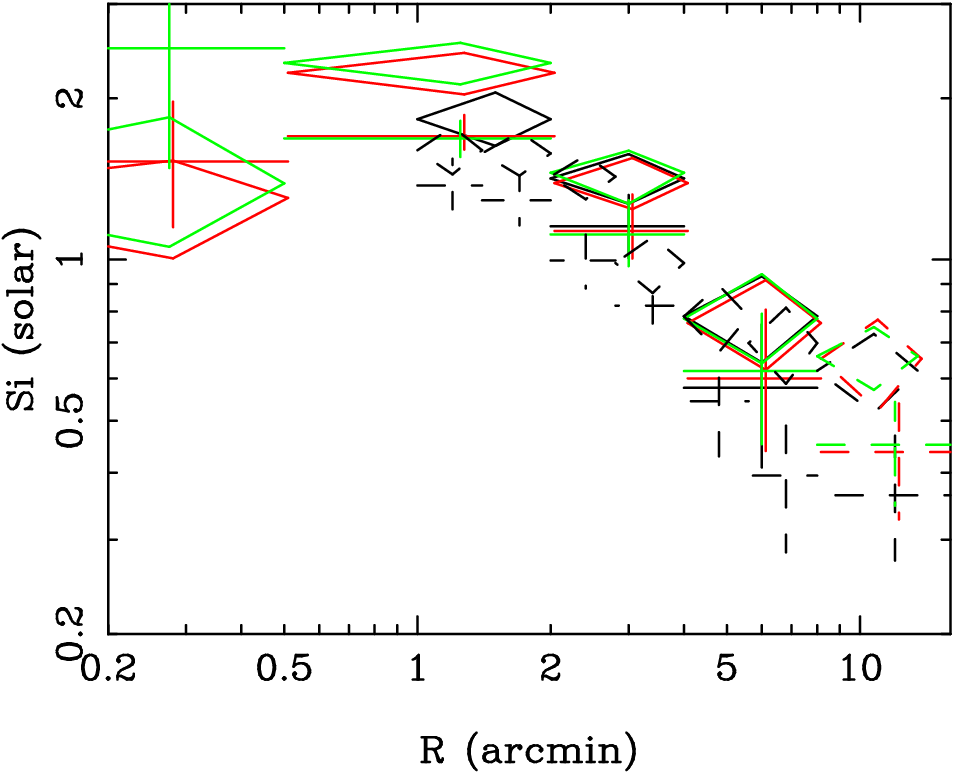}}
\resizebox{8.1cm}{!}{\includegraphics{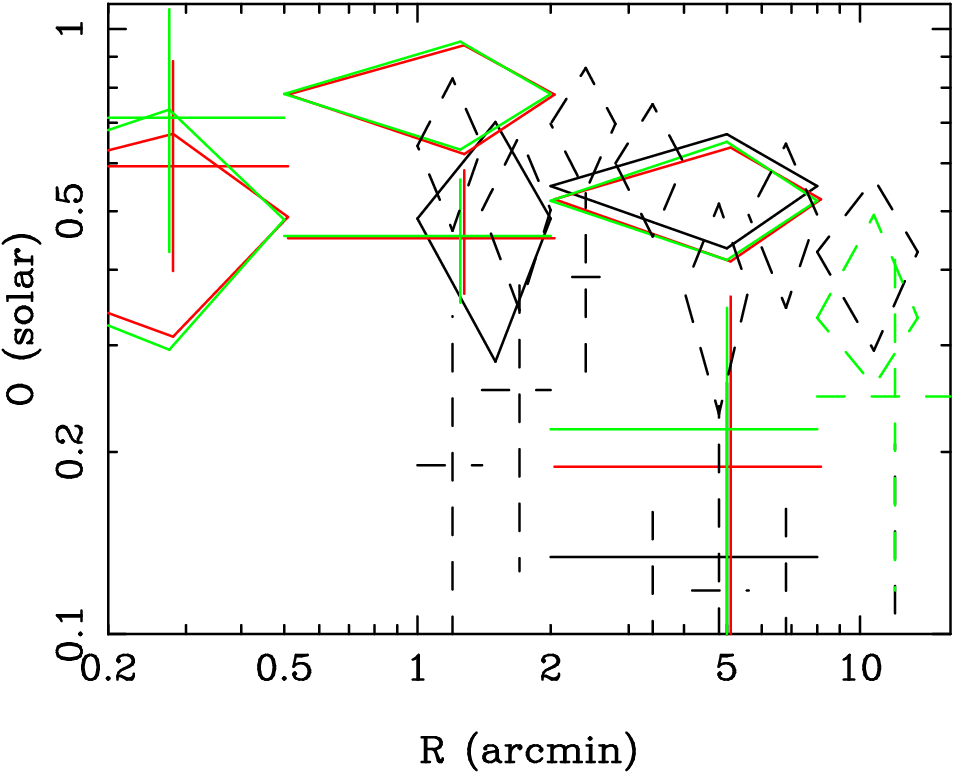}}
\caption{Radial profiles of the abundances of Fe, Si and O.
The meanings of symbols are the same as those of Figure 2.
The data with a reduced $\chi^2>2$ are not shown.
}
\label{abund}
\end{figure}

The temperature of the ICM decreases  sharply toward the center
(Figure \ref{ktnh}), and
 we need more than two temperature components to fit the deprojected
spectra at the central region.
The observed sharp temperature gradient means that
the single temperature
model is not enough to fit each deprojected spectrum,
even if the ICM is isothermal locally.
In addition, the cluster is not circularly symmetric and
the  cool plume-like structure at the center (Sanders \&
Fabian 2002) might introduce an additional temperature
component.

 The hydrogen column density, ${\rm {N_H}}$ is  consistent with the
Galactic value  at $R>1'$,  although the EPN and the
EMOS give slightly different results by a few times $10^{20}{\rm
cm^{-2}}$ (Figure \ref{ktnh}).
Within $R<1'$,  the two- and the multi-temperature models give similar
values to those of $R>1'$.
Higher column densities derived from the single
temperature model fit within $R<1'$ are not acceptable due to their high $\chi^2$.
Therefore,  ${\rm {N_H}}$ is consistent with the Galactic value within the whole field of view.

 Discarding the results  of  unacceptable fits (reduced $\chi^2>2$), 
 the  Fe abundance is $\sim$0.6 solar  at $R>8'$ and 
  within the radius, it increases steeply toward center.
Within $2'$, the gradient becomes weaker.
The average value of the Fe abundance at $R<2'$ is 2.3 solar  (Figure \ref{abund}).
The EPN and the EMOS give  consistent Fe abundances within 10\% (Table 2).
The Fe abundances from  the deprojected spectra are slightly
higher than those from the projected spectra.

\begin{figure}[]
\centerline{\resizebox{8.1cm}{!}{\includegraphics{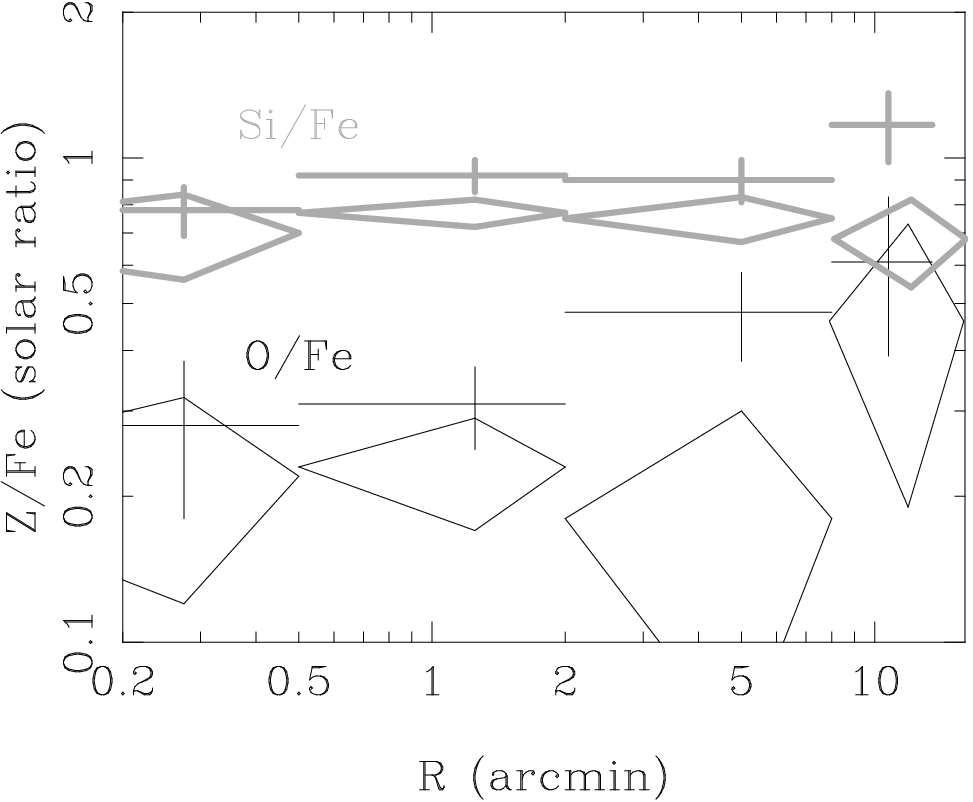}}}
\caption{Radial profiles of the O/Fe (black lines) and the Si/Fe
 (gray lines)  ratios in units of the
 solar elemental ratio derived from the multi-temperature MEKAL model of  the EMOS
 (diamonds) and the EPN (crosses) spectra.
}
\label{abund_ratio}
\end{figure}

 At each radius, the Si abundance 
 is close to the Fe abundance (Figure \ref{abund}, \ref{abund_ratio}).
The uncertainties in the Si/Fe ratio are better constrained than those in the Si and
 Fe abundances themselves (\ref{abund_ratio}), since the derived abundance of  Si and Fe
are correlated to each other.
On average, the Si/Fe ratio is 0.89$\pm$0.04  and 0.75$\pm$0.04 in units of the solar
 ratio using the EMOS and EPN, respectively.

 The O abundances are much smaller than the Si and Fe abundances
(Figure \ref{abund}, 4).
 The two detectors give inconsistent O abundances.
Using the EMOS, 
the O abundance is about 0.8 solar at $R$=0.5--2.0$'$ and at $R>$2$'$,
it reduces to 0.4 solar (Figure \ref{abund}),
and the O/Fe ratio increases toward outer regions (Figure \ref{abund_ratio}).
In contrast, the EPN gives a significantly  smaller O
abundance than the EMOS especially in the outer regions (Figure \ref{abund}),
and the profile of the  O/Fe ratio is consistent with no radial gradient
(Figure \ref{abund_ratio}).

In summary, as in center of other clusters and groups of galaxies,
the Fe and Si abundances are close to  each other with strong negative radial
gradients.
The O abundance is at least a factor of 2 smaller.
 The discrepancies in the O and Si abundances between the EMOS and
EPN will be discussed in section 4.3.
\section{Uncertainties in the abundance determination}

 In the abundance determination, we must examine uncertainties such as the
dependence on the plasma code and on the background subtraction.
In this section, we study these uncertainties and also calibrate the
abundances with the strengths of the lines.

\subsection{The effect of the plasma code}

In order to constrain the effect of the plasma code on the abundance determination, 
  we compared the results from the APEC (Smith et al. 2001) model with those
from the MEKAL model.
We  used the APEC version 1.1  model in XSPEC version 11.2.
As in the MEKAL model fit, we fitted the deprojected spectra with a single,
and  a multi-temperature APEC model.

\begin{figure}[]
\resizebox{8.1cm}{!}{\includegraphics{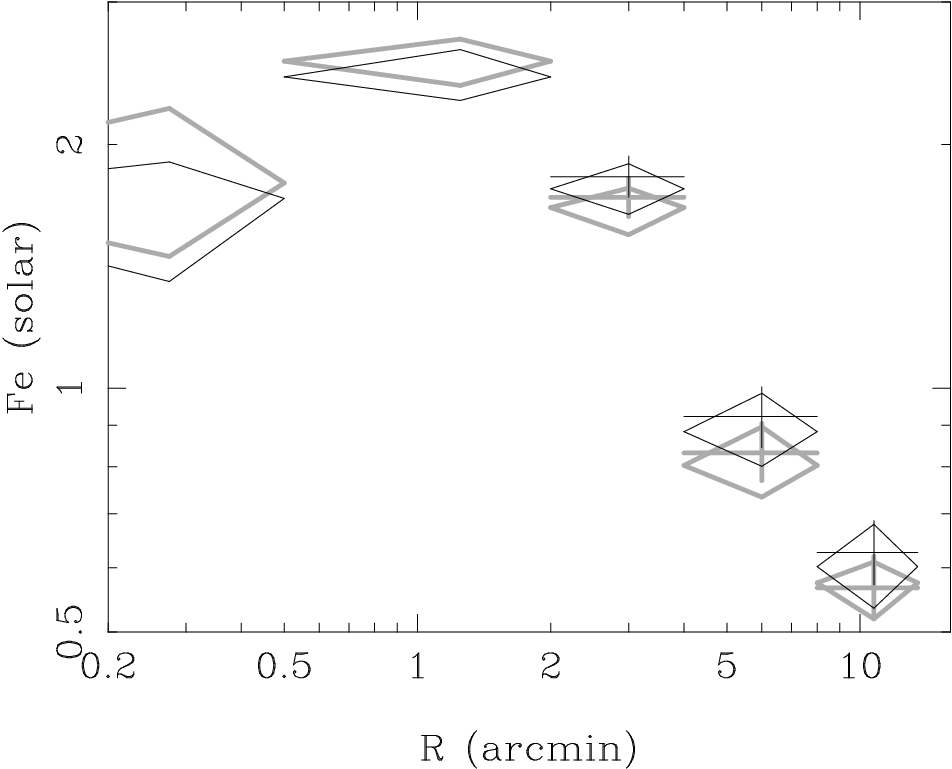}}
\caption{The Fe abundance profile derived from the MEKAL (gray data) and the APEC
 (black data) model, from the single-temperature model (crosses)
 and the multi-temperature model (diamonds).The data with a reduced $\chi^2>2$ are not shown.}
\label{apec_fe}
\end{figure}

\begin{figure}[]
\resizebox{8.1cm}{!}{\includegraphics{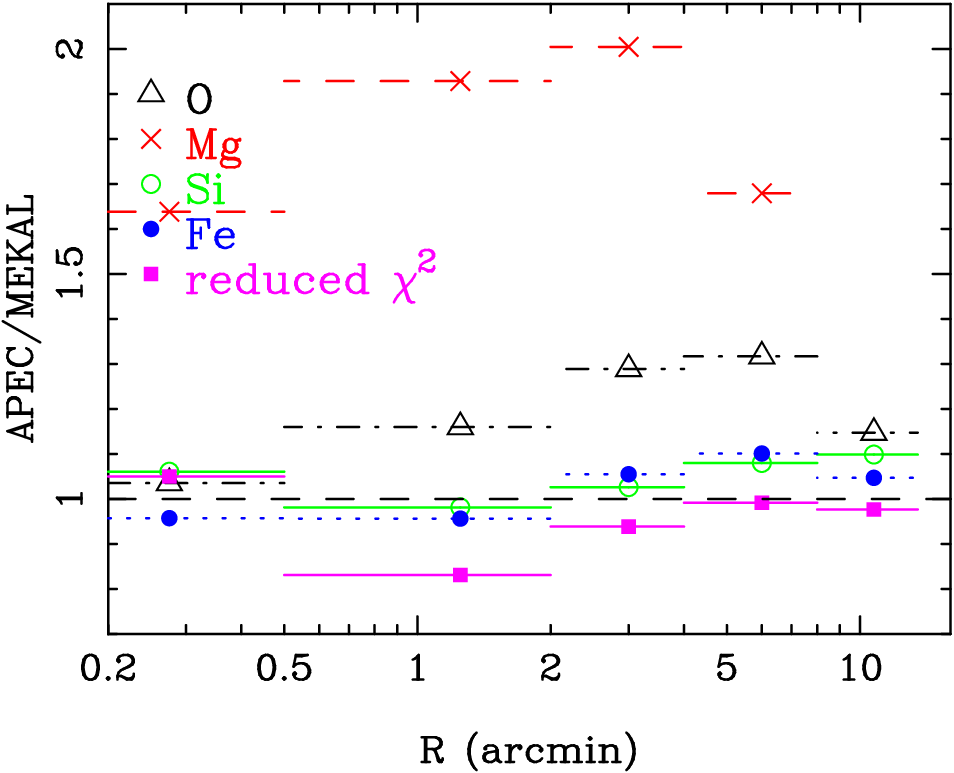}}
\caption{ The ratio of the O(open triangles), Mg(crosses), Si(open
 circles) and Fe(closed circles) abundances and the reduced $\chi^2$
(closed boxes) between the multi-temperature APEC and the multi-temperature MEKAL model.
}
\label{apec_all}
\end{figure}

The results are summarized in Table 3.
With the APEC model, the fit to the Fe-L/Mg-K region at 1.2--1.5 keV
improved  from that with the MEKAL code (Figure 1).
As a result, the reduced $\chi^2$ are consistent or  slightly smaller
than those from the MEKAL model fit.
The derived temperatures, hydrogen column densities, and
the  Si and Fe abundances are consistent within 10\% with those
from the MEKAL code (Figure \ref{apec_fe}, \ref{apec_all}).

However, the O and the Mg abundances increased by $\sim$20\% and$\sim80$\%,
 respectively,   from the MEKAL model fit.
The difference in the O abundance is consistent with the difference of the 
 strength of the H-like O line between the APEC and MEKAL models,
since at 2 keV, that of APEC is 10\% larger than that of MEKAL model,
and at 4 keV, the difference increases to 30\%.
The difference in the  Mg abundance is not due to the difference in 
the strength of the Mg-K line between the two codes  which is smaller than that of O below 4 keV.
It should be caused by the difference in the Fe-L modeling around the Mg-K line.
We will discuss it later in \S4.3.2.

\begin{table*}
\caption{Results of spectrum fitting of the deprojected spectra within
 8$'$ and projected spectrum at 8--13.5$'$ of the EMOS with the
 single- and the multi-temperature APEC model.}
 \begin{tabular}[t]{lllllllllr}
 $R$     &  &  & kT    & $N_{\rm H}$ & O & Mg & Si  & Fe &  $\chi^2/dof$\\
(arcmin) & &&(keV)   & (10$^{20}{\rm cm^{-2}}$) & (solar)&
  (solar)& (solar)& (solar)& \\
 \hline
 0.00- 0.50 &EMOS &1T &1.00 &17.1& 0.36& 0.23& 0.39& 0.27& 676/129 \\
 0.00- 0.50 &EMOS &14T& & 10.6$^{+2.7}_{-1.5}$& 0.50$^{+0.23}_{-0.21}$& 1.45$^{+0.49}_{-0.49}$& 1.47$^{+0.34}_{-0.40}$& 1.72$^{+0.19}_{-0.36}$& 146/116 \\
\hline
 0.50- 2.00 &EMOS &1T &1.93 &11.1& 0.47& 0.68& 1.35& 1.61& 585/249 \\
 0.50- 2.00 &EMOS &14T&  &9.3$^{+0.6}_{-0.7}$& 0.91$^{+0.21}_{-0.19}$& 1.54$^{+0.15}_{-0.24}$& 2.29$^{+0.25}_{-0.20}$& 2.42$^{+0.20}_{-0.16}$& 266/236 \\
\hline
 2.00- 4.00 &EMOS &1T &2.94$^{+0.06}_{-0.06}$ &9.8$^{+0.6}_{-0.5}$&
  0.96$^{+0.27}_{-0.25}$& 0.89$^{+0.28}_{-0.27}$&
  1.46$^{+0.17}_{-0.16}$& 1.83$^{+0.11}_{-0.10}$& 346/294 \\
 2.00- 4.00 &EMOS &14T&  &9.5$^{+0.7}_{-0.6}$& 1.00$^{+0.26}_{-0.25}$& 0.89$^{+0.28}_{-0.25}$& 1.49$^{+0.19}_{-0.16}$& 1.76$^{+0.13}_{-0.12}$& 301/281 \\
\hline
 4.00- 8.00 &EMOS &1T &3.53$^{+0.11}_{-0.11}$ &9.1$^{+0.6}_{-0.5}$& 0.61$^{+0.27}_{-0.26}$& 0.62$^{+0.31}_{-0.30}$& 0.84$^{+0.16}_{-0.16}$& 0.92$^{+0.08}_{-0.08}$& 397/332 \\
 4.00- 8.00 &EMOS &14T&  &9.1$^{+0.5}_{-0.7}$& 0.54$^{+0.27}_{-0.23}$& 0.58$^{+0.32}_{-0.28}$& 0.84$^{+0.18}_{-0.15}$& 0.88$^{+0.10}_{-0.08}$& 372/319 \\
\hline
 8.00-13.50 &EMOS &1T &3.77$^{+0.10}_{-0.10}$ &7.4$^{+0.4}_{-0.4}$&  0.52$^{+0.18}_{-0.18}$& & 0.67$^{+0.12}_{-0.11}$& 0.63$^{+0.06}_{-0.06}$& 443/325 \\
 8.00-13.50 &EMOS &14T& & 8.7$^{+1.1}_{-1.1}$& 0.38$^{+0.18}_{-0.16}$& & 0.72$^{+0.13}_{-0.12}$& 0.60$^{+0.08}_{-0.07}$& 388/312 \\
\hline
\end{tabular}

\end{table*}


\subsection{Effect of the  background uncertainty}

\begin{figure}[]
\resizebox{8.1cm}{!}{\includegraphics{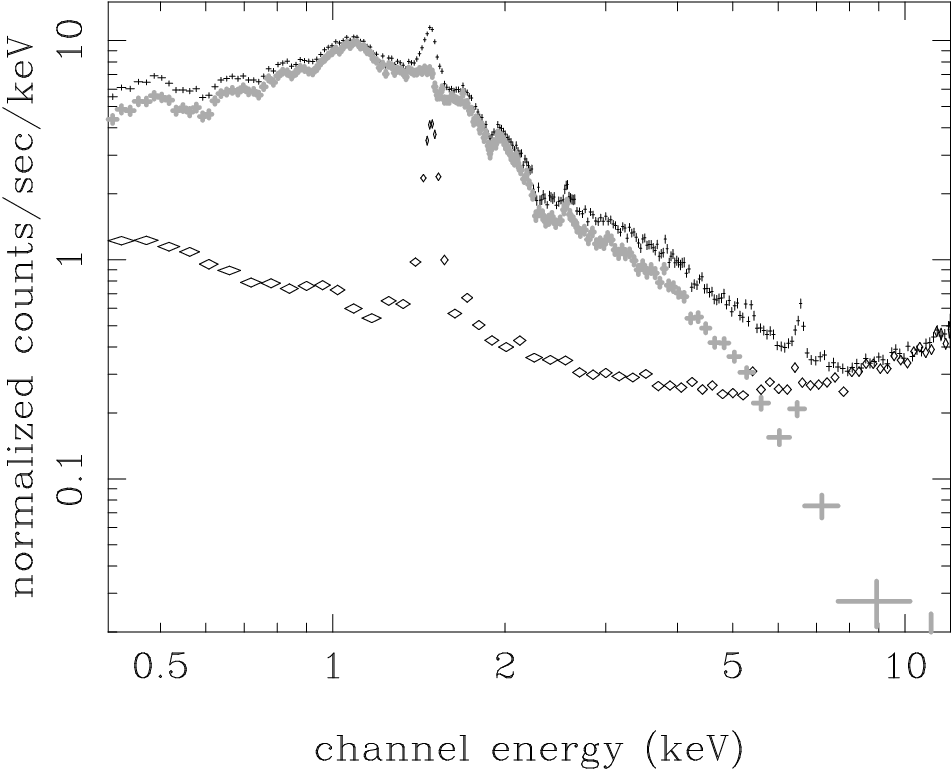}}
\caption{The raw EMOS spectrum at $r>8'$ (black crosses) and 
the adopted background spectrum (black diamonds), and the background
 subtracted spectrum (gray crosses).
}
\label{mos_spec_bgd}
\end{figure}

 One of the most severe problems in analyzing  XMM-Newton data 
are uncertainties in the background.
In this paper, we adopted  deep survey data as a background, which
agree well with the data of the Centaurus cluster above 8 keV (Figure \ref{mos_spec_bgd}).
In addition, in order to study the effect of the uncertainty of the background, we
artificially scaled  the background by 10\%, and 
fitted the projected EMOS spectra at $r>2'$ in the same way as previously.
Since spectral variations of the  background amount to about 10\% below 1 keV
when the background is scaled with the count-rate outside the field of
the view of the detector (Katayama et al. 2004), and
the relatively high  Galactic $N_{\rm H}$
towards the Centaurus cluster absorbs the low energy photons from the
Cosmic X-ray Background (CXB).

\begin{figure}[]
\resizebox{8.1cm}{!}{\includegraphics{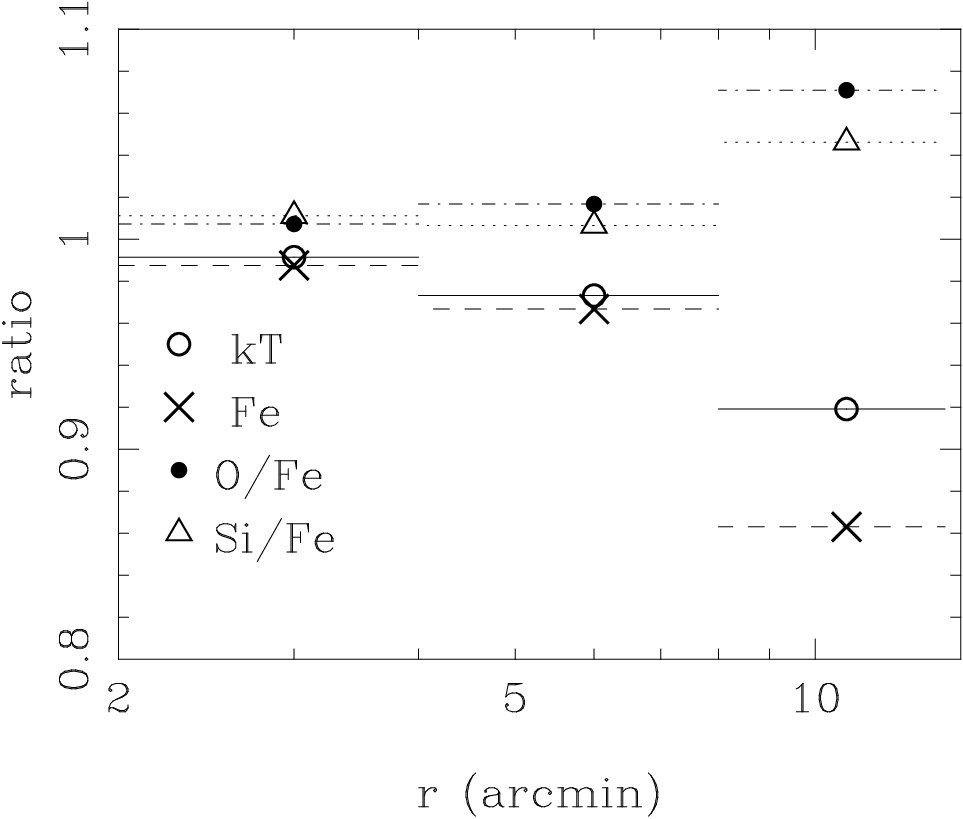}}
\caption{
The ratio of derived parameters of the projected EMOS spectra with the
 multi-temperature  model using a  background scaled by a factor of 1.1
to those using the unscaled background used in the paper.
}
\label{bgd}
\end{figure}

The results are summarized in Figure \ref{bgd}.
Within 8$'$, the derived temperatures, the Fe abundances, the O/Fe and
the Si/Fe ratios are consistent within a few \% with those derived from
the background used in \S3.1.
Outside 8$'$, the 10\% change of the background changes derived
parameters by $\sim$ 10\%.
Therefore,  the background problem is not severe in this observation.

\subsection{Comparisons between the line strength and model fit}
\subsubsection{The line strengths of O and Si and the abundance determination}

In order to study the discrepancy in the O and the Si abundances
between the EMOS and the EPN data,  it is important to check their line strengths directly.
Complications arise
since the energy resolution of the EMOS has become worse due to the 
radiation damage in orbit (Kirsh 2002),
 and that of the EPN depends on the distance from the
read-out. 
In addition, there are still discrepancies of several \% in the 
response matrix between the EPN and EMOS,
 which might give a large uncertainty in strengths of weak lines.

\begin{figure}
 \centerline{
\resizebox{4.1cm}{5.5cm}{\includegraphics{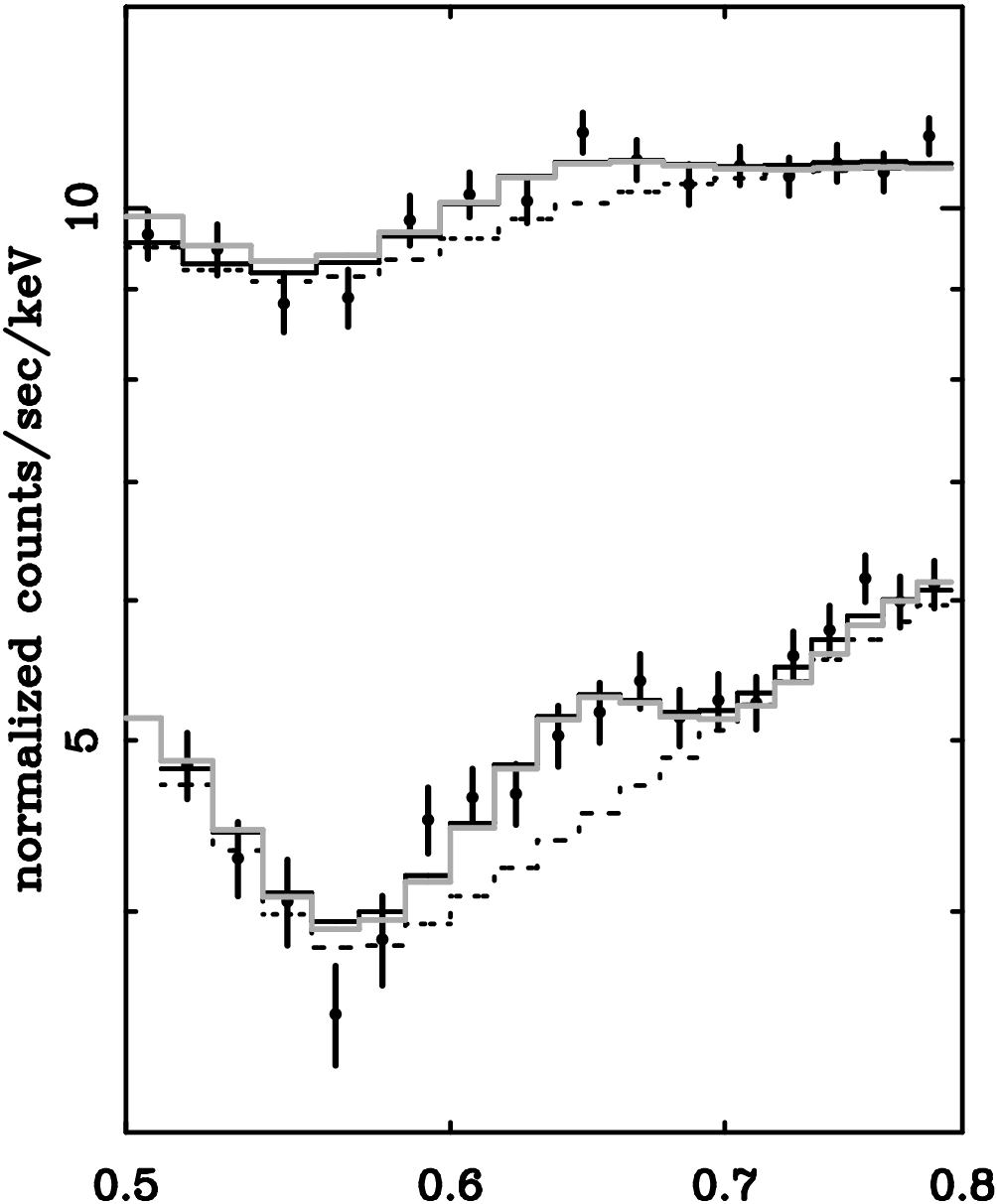}}
\resizebox{3.9cm}{5.5cm}{\includegraphics{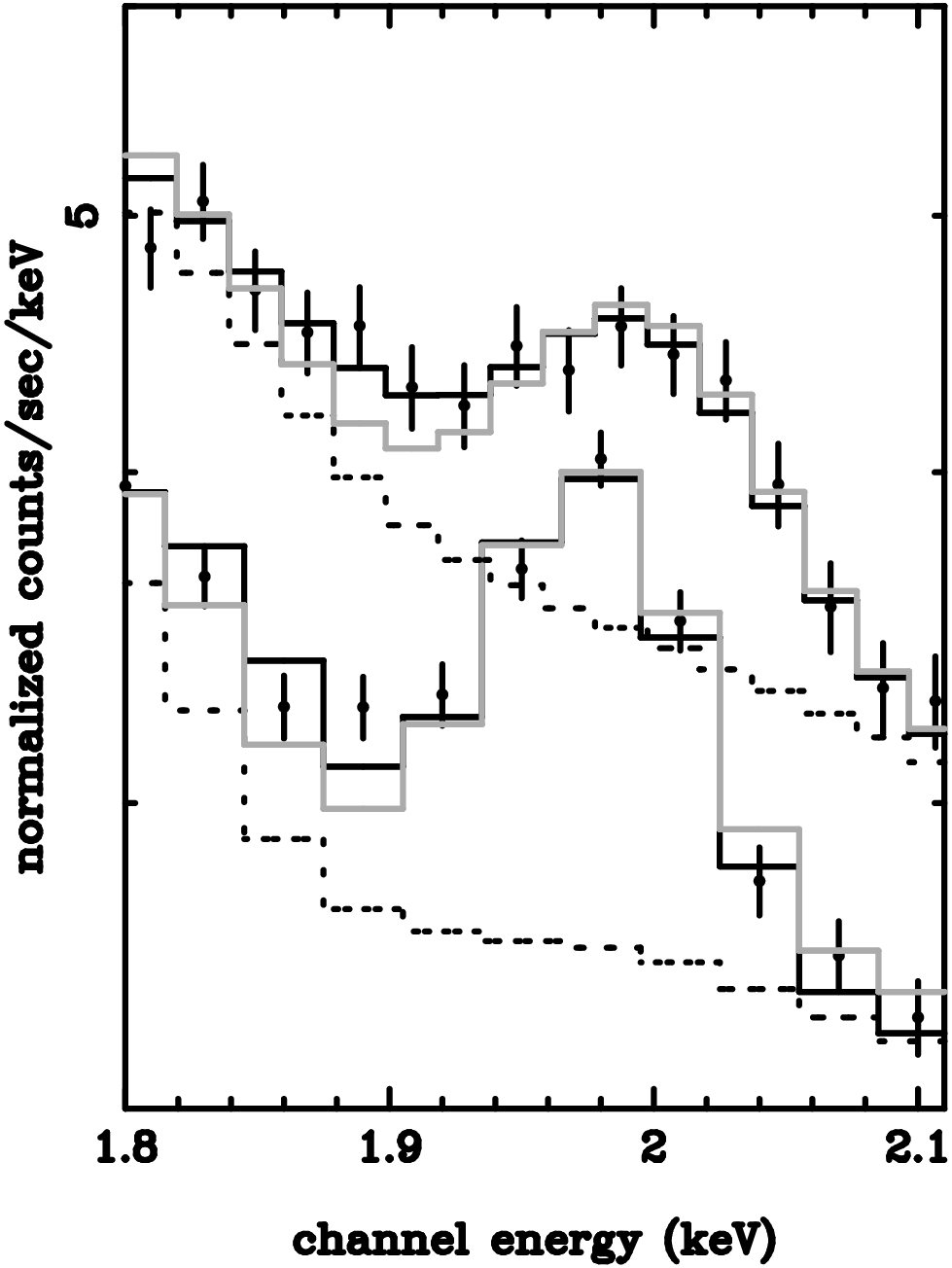}}
}
\caption{
Deprojected spectra of the EMOS (EMOS1+EMOS2; lower curve) and the
 EPN (upper curve) at $R$=2.0--8.0$'$  around
the energy of  the lines of O (left panel) and Si (right panel).
The gray data points and black solid lines
 correspond to the best fit model with one temperature MEKAL model and
that with Gaussians with 0 line widths, respectively.  The black dotted lines represent
the continuum emissions as described by the Gaussian fit.
}
\label{osiline}
\end{figure}

\begin{figure}[]
\centerline{
\resizebox{8.1cm}{6cm}{\includegraphics{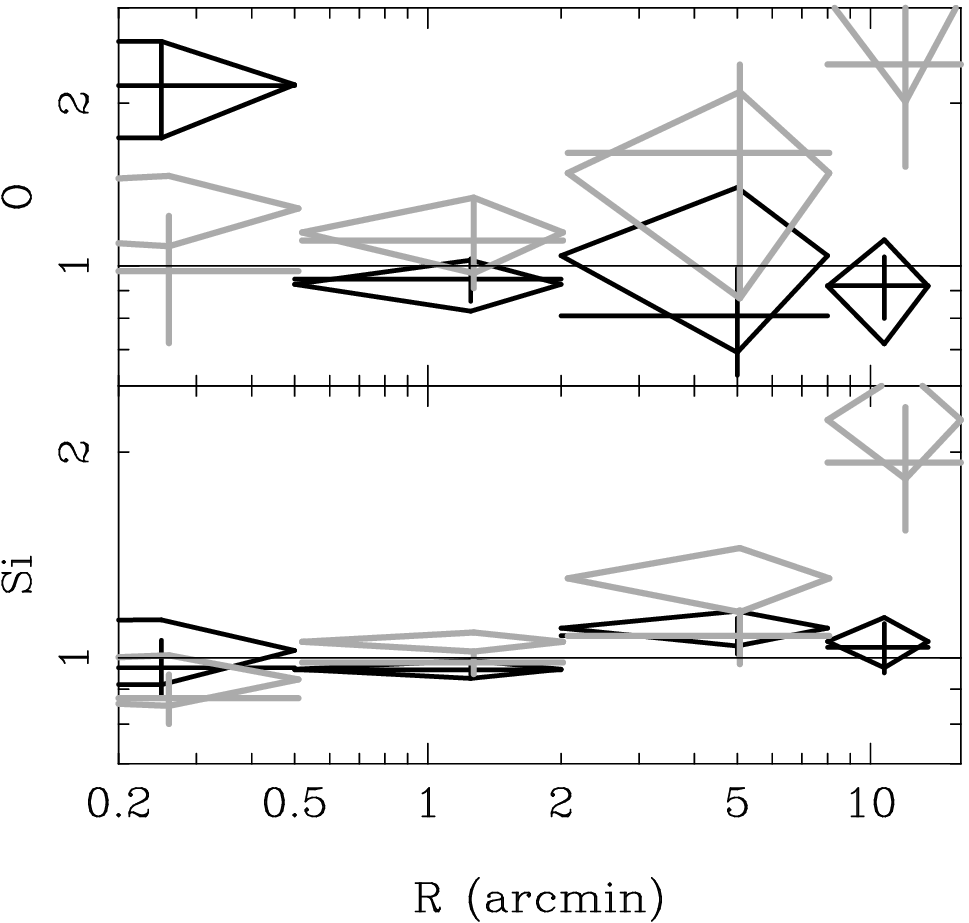}}
}
\caption{The ratio of line strengths K$\alpha$ of H-like O (upper panel) and Si (lower panel)
 derived from the Gaussian fit to those of best fit values from spectral fit for the EMOS (black lines)
and EPN (gray lines).
Crosses and diamonds correspond to the fitting with a Gaussian with 0 line width and free line width, respectively.
Errors correspond the 68\% level.
}
\label{eqw}
\end{figure}

The  strengths of K$\alpha$ lines of H-like O and Si are derived
 from fittings of the spectra around the lines with Gaussians. The details are
the same as in Matsushita et al. (2003).
First, the line widths of the Gaussians were frozen to  zero, and then
allowed to be free.
Figure \ref{osiline} compares the best fit 
MEKAL model and the Gaussian fits for  representative spectra at
$R$=2--8'.
In the EMOS spectrum, we can see a weak bump of the H-like O line, and a clear
bump of the H-like Si line.
In contrast, in the spectrum of the EPN, which has an  energy resolution
 worse than that of
the EMOS, the continuum and the O line are hardly
distinguished and the  Si line is much broader than the EMOS one.

For the EMOS spectra, the line strengths of K$_\alpha$ of H-like O and
Si from the Gaussian fits are consistent within 10\% 
with the those from the MEKAL model, except for the O line strength within 
0.5$'$, where the O line is completely hidden in the Fe-L lines (Figure \ref{eqw}).
This result indicates that the effect of the decrease of the energy resolution of the EMOS detector
should not affect  the derived O and Si abundances.

In contrast,
allowing the line width to be a free parameter,
the Gaussian fits to the EPN spectra give systematically larger line
strengths than the MEKAL model fit.
As a result, adopting the values from the Gaussian fits,  
the EPN abundances become consistent with the EMOS ones.
 Therefore,
the O and Si abundances derived from the EMOS fit should be reliable,
since it has a better spectral resolution and it gives consistent 
 strengths of the K$\alpha$ lines between the Gaussian fits and the
 MEKAL model fits.

\begin{figure}[]
\centerline{
\resizebox{8.1cm}{6cm}{\includegraphics{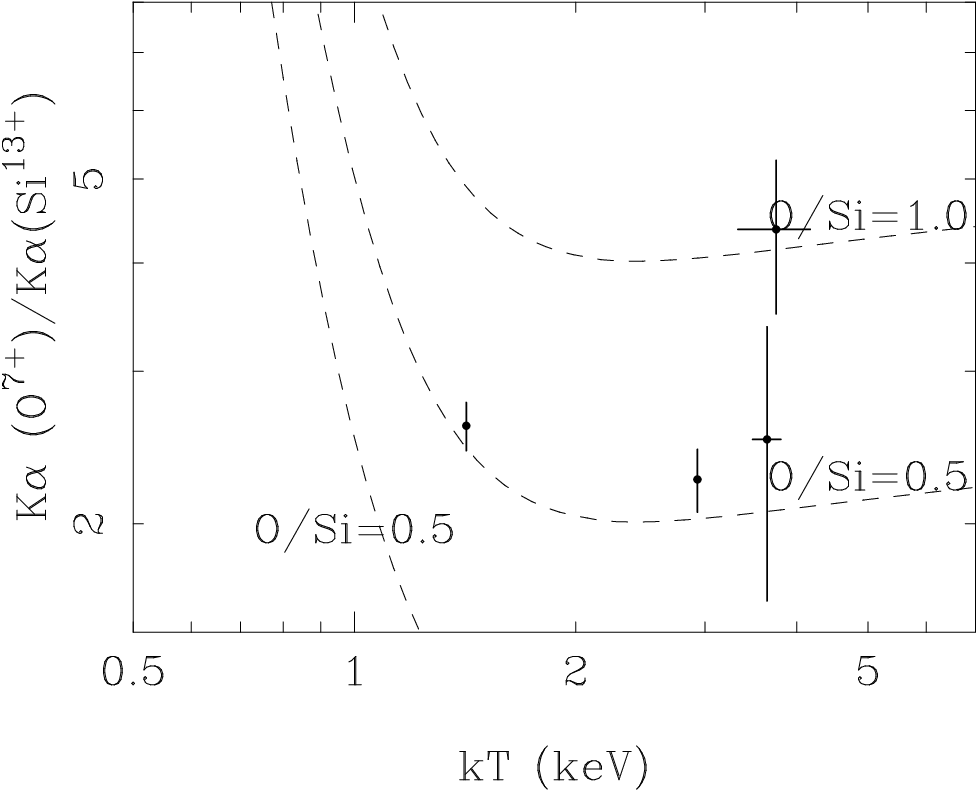}}
}
\caption{The line ratios of K$\alpha$ lines of H-like O and Si of the 
projected spectra of the EMOS, plotted
 against  the best fit temperature fitted with the single MEKAL model.
Dashed lines  correspond to constant abundance
 ratios in units of the solar ratio.
Errors represent the 68\% confidence level.
}
\label{osiratio}
\end{figure}

As shown in the Figure \ref{osiratio},
the ratio of strengths of K$\alpha$ lines of H-like O and Si becomes constant above
1.7 keV and increases toward lower energies.
This temperature dependence means that
any temperature distribution cannot yield a  O/Si ratio larger than 0.6 solar within 4$'$.
Furthermore, we got a higher value of the line ratio at $R>8'$ by a factor of
2.
It implies that the O/Si ratio increases  in the region, since
the contribution to the O line from the component below 1.7 keV of the
multi-temperature MEKAL model is smaller than 10\%.

\subsubsection{The line strength of the Mg-K}

It is also important to  derive the strengths of the K$\alpha$-lines of
Mg  directly, since the Mg abundances
derived from the spectral fits depend on the modeling of the underlying Fe-L
lines. We therefore fitted the EMOS spectra within  the  energy range of 1.3--1.65 keV
with Gaussians plus one- two- or three-temperature MEKAL or APEC model.

\begin{figure}[]
\centerline{
\resizebox{8.1cm}{6cm}{\includegraphics{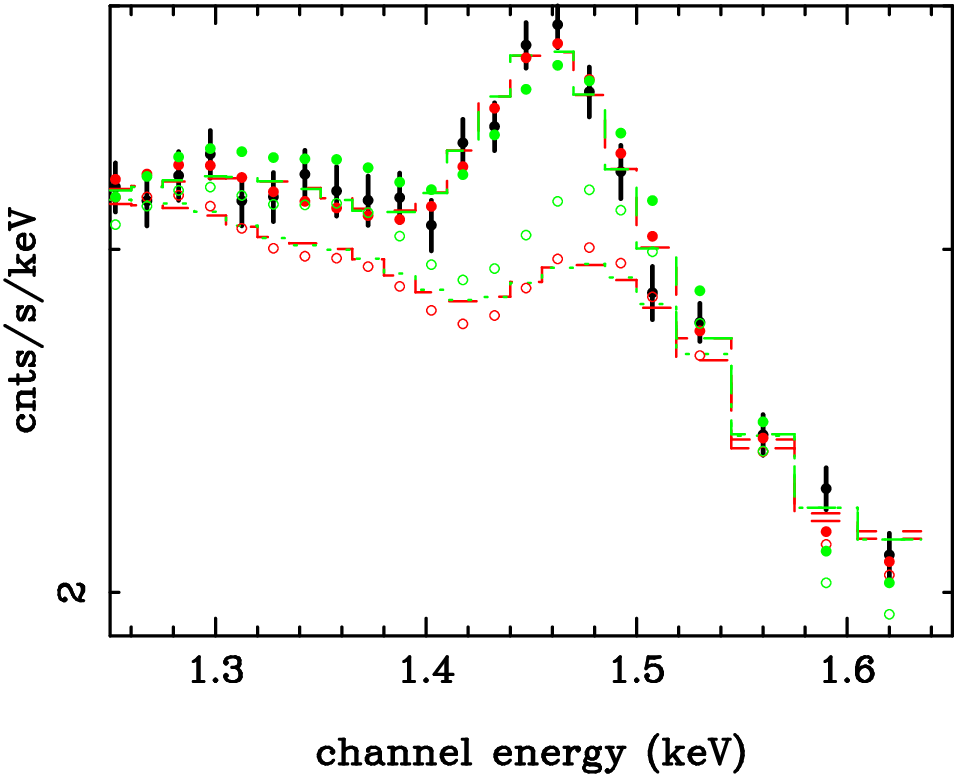}}
}
\caption{
The deprojected  spectrum  of the EMOS at $R$=0.5--2.0$'$ (black closed circles) 
fitted with the multi-temperature MEKAL model (green closed circles),
the multi-temperature APEC model (red closed circles),
the Gaussians with MEKAL models (green dashed lines) and the Gaussians
with the APEC models (red dashed lines).
The contributions of the continuum plus the Fe-L are also shown for the
multi-temperature APEC model (red open circles) and the
 multi-temperature MEKAL model (green open circles) and the Gaussian fit
(red and green dashed lines).
}
\label{mgkspec}
\end{figure}

\begin{figure}[]
\centerline{
\resizebox{8.1cm}{6cm}{\includegraphics{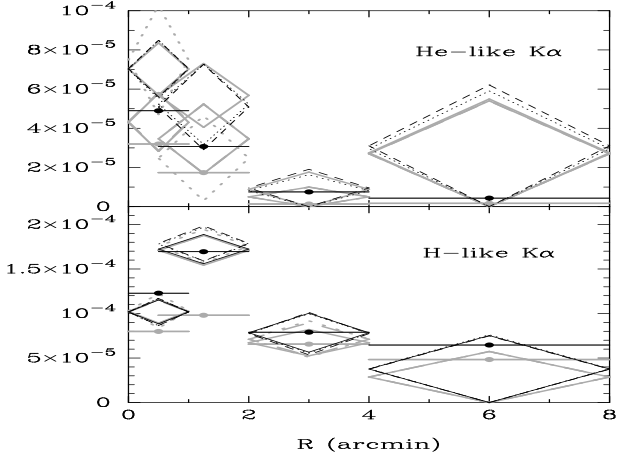}}
}
\caption{
The profile of the line strengths of the K$\alpha$-line of He-like 
and H-like Mg of the best fit model of the multi-temperature model (horizontal solid
 lines) and derived from the Gaussians plus a single temperature model
(solid diamonds), two temperature model (dashed diamonds), and
 three-temperature model (dotted diamonds).
The gray and black data correspond to the MEKAL and APEC model,
 respectively.
Errors correspond the  68\% confidence level.
}
\label{mgline}
\end{figure}

 All of  the Gaussian models can fit the spectra around the
Mg-K/Fe-L  region well (Figure \ref{mgkspec}). 
The derived  strengths of the K$\alpha$-line of H-like and He-like Mg 
do not depend on the temperature structure (i.e. one-, two, or
three-temperature model) or the  plasma code (i.e. MEKAL or APEC
model). 
The line strengths agree well with those  derived from the fitting
the spectra
 described in \S4.1, but are systematically higher than those derived
from the multi-temperature MEKAL model described in \S3,
which cannot reproduce the spectra around the Mg-K/Fe-L region (Figure \ref{mgline}).
Therefore, it is reasonable to adopt the Mg abundances derived
from the multi-temperature APEC model fit as the Mg abundances of the ICM.

 We  note that  the Mg/O ratio  does not depend on the temperature structure 
since the line ratio of K$\alpha$ lines of these two elements is nearly
 constant above 1.1 keV (Matsushita et al. 2003).

\subsubsection{The line strength of Fe-K}\label{FeK}

\begin{figure}[]
\centerline{
\resizebox{8.1cm}{6cm}{\includegraphics{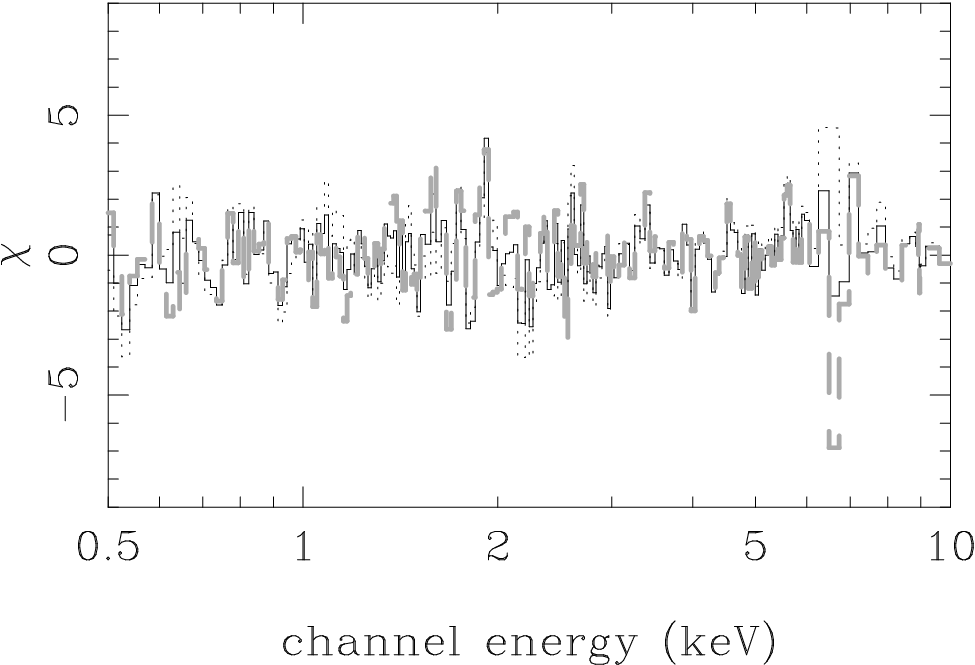}}
}
\caption{The  $\chi$ distributions of the deprojected spectrum at
 $R$=0.5--2.0$'$ fitted with the multi-temperature APEC model. 
The solid black line corresponds to the best fit model, and the dotted black
 line and dashed gray lines correspond to the model where the Fe
 abundance is fixed to be 1.6 solar and 3.2 solar, respectively.
}
\label{fechi}
\end{figure}

Since the temperature dependence  of the intensities of the Fe-L and Fe-K lines
is completely different,  a comparison between the Fe-L and the Fe-K
lines  strongly constrains
the temperature structure and the Fe abundance.
In the spectral fitting by the MEKAL  model or the APEC model, 
the strengths of the 6.7 keV Fe-K lines are fitted well, although the
fit to the Fe-L lines mostly determines the Fe abundance due to its high
statistics (Figure \ref{fechi}).

When the Fe abundance is shifted by 40\% in the fits
 with the multi-temperature model,  
the Fe-L structure can  be roughly fitted  although the $\chi^2$
increased significantly (Figure \ref{fechi}).
However,
the strengths of the Fe-K line cannot be reproduced.
Therefore, the uncertainties in the Fe-L line fit, due to the temperature structure and
the plasma code should not be  severe problems at least in the Centaurus cluster.

\begin{figure}[]
\centerline{
\resizebox{8.1cm}{6cm}{\includegraphics{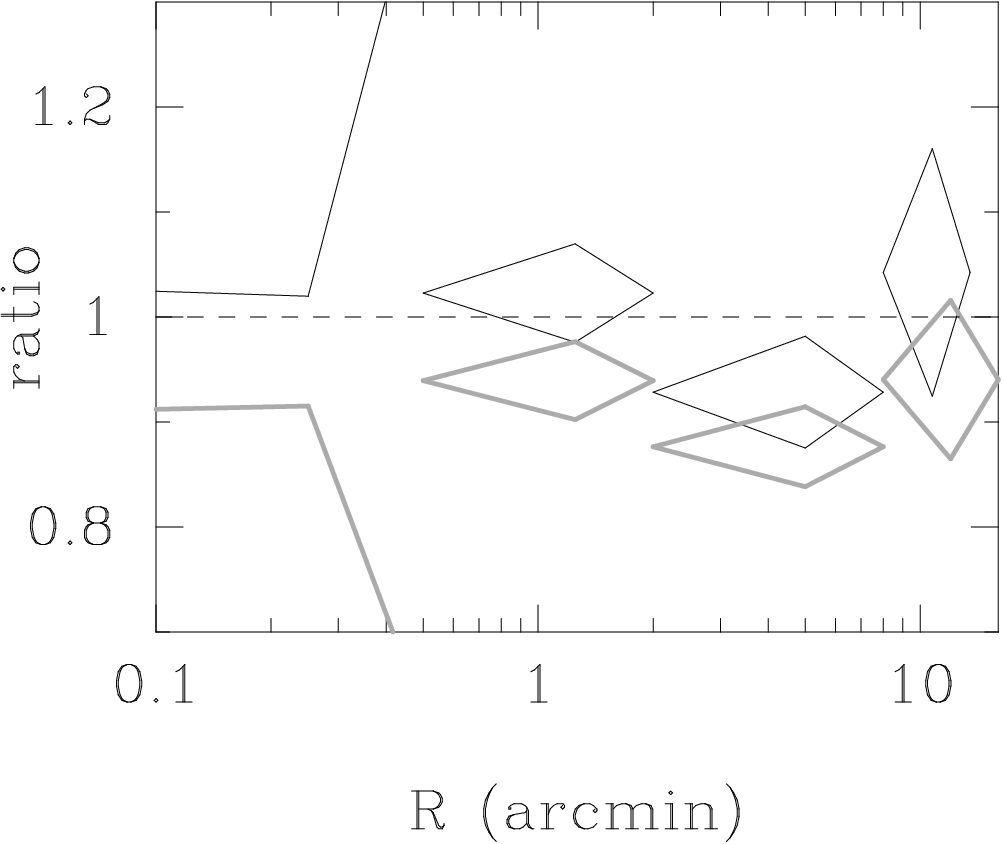}}
}
\caption{ The ratio of line strengths of the 6.7 keV Fe-K line
 derived from the Gaussian fit to those of best fit values from spectral
 fit for the EMOS (black lines) and EPN (gray lines), where the Fe-L is
 important to determine the Fe abundance.
Errors correspond to the 68\% level.
}
\label{fek}
\end{figure}

Since abundances derived from the Fe-K line may have an uncertainty
due to the uncertainty of the background, 
we have also derived the line strength of the Fe-K line directly from a
Gaussian fit. 
The width of the Gaussian was allowed to be a free parameter,
since the 6.7 keV Fe-K consists of several strong lines.
In figure \ref{fek}, the derived line strengths from the Gaussian fit are compared to the 
line strengths of the Fe-K line derived from the best fit
multi-temperature MEKAL model. 
These values agree  within 10\%.
Thus,  the line strengths of the Fe-K line are consistent with the Fe
abundances derived from the MEKAL  model fit and a change of the Fe
abundance cannot reproduce the  observed line strength of the Fe-K.

\section{Discussion}

\subsection{Summary of the results}

The abundances of O, Mg, Si, and Fe in the ICM of the Centaurus cluster are derived.
  We have checked uncertainties in the temperature structure, 
 in the background subtraction, and in the spectral model and
studied their effects on the abundance determination.
The Fe abundance is $\sim2.3$ solar within 2$'$ and decreases to 
$\sim$0.6 solar at $R>8'$.
The Si abundance is close to the Fe abundance.
In contrast, the  O and Mg abundances  are  much  smaller than 
the Fe and Si abundances in units of the solar ratio.

The abundance pattern of the ICM of the Centaurus cluster, the low O abundance compared to those
of  Si and Fe, resembles to those observed in center of other clusters
and groups of galaxies, such as M 87 (Matsushita et al. 2003), A496 (Tamura et al. 2001b),
Fornax cluster (Buote et al. 2002) and
group-center elliptical galaxies NGC 4636 (Xu et al. 2002), NGC 5044 (Tamura
et al. 2003;  Boute et al. 2003).
However, there are small differences in the observed abundance pattern of the ICM.
For example, 
the central Fe abundance of the Centaurus cluster (2.3 solar) is 40\%
larger than
that of M 87 (1.6 solar). In addition, the Si/Fe ratio of the former is 20\% smaller than that
of the latter.
 In this section, we mainly compare and discuss the abundance pattern of the Centaurus 
clusters with that of M 87, since
these two clusters are the nearest and the best studied objects among
relatively low temperature clusters.

\subsection{Comparison with stellar metallicity and SN II abundance pattern}
The O and Mg in the gas around a cD galaxy is a mixture of those in the
ICM and those in the accumulation of  stellar mass loss from the cD
galaxy, since these elements are only synthesized by SN II.
The O and Mg abundances of the mass loss  reflect those of mass losing stars,
which contribute significantly to the optical light from elliptical galaxies.
The stellar metallicity of elliptical galaxies are usually derived
from the Mg$_2$ index in the optical spectra,
which  depends mainly on the Mg abundance and stellar metallicity 
where O contributes most, and weakly on the age population of stars.
Therefore, comparing the O and Mg abundances of the gas and stellar
metallicity derived from the optical Mg$_2$ index is important to study 
the gas flow and contribution of stellar mass loss, and stellar metallicity
of the cD galaxies.

\begin{figure}[]
\centerline{
\resizebox{8.1cm}{6cm}{\includegraphics{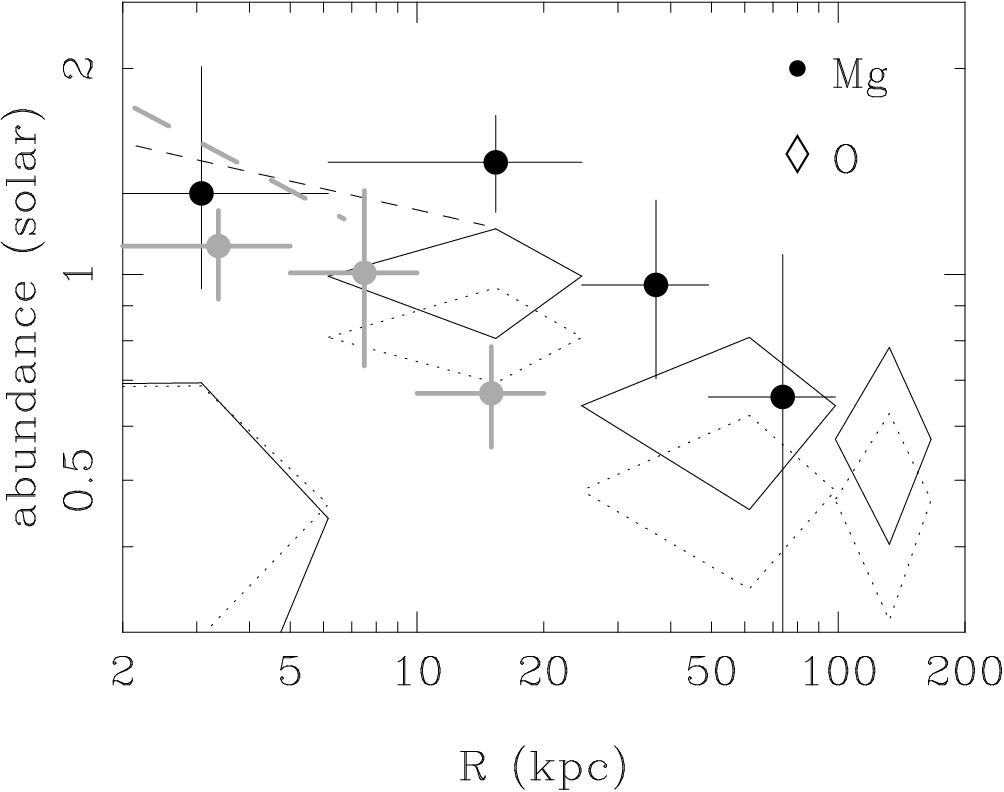}}}
\caption{ Mg (closed circles) and O (diamonds) abundances derived
for the Centaurus cluster (black) and M 87 (gray)
 from the multi-temperature MEKAL (dotted lines) and APEC (solid lines)  model.
The dashed lines represent the stellar metallicity of  NGC 4696
 (black) and M 87 (gray)  derived from the optical Mg$_2$ index (Kobayashi \& Arimoto 1999).
}
\label{mgo}
\end{figure}

The Mg abundances of the two clusters are close to  their stellar metallicity derived
from the Mg$_2$ index  (Figure \ref{mgo}; Kobayashi \& Arimoto 1999).
 The gradient of the Mg abundance of the gas around M 87 is steeper than in the Centaurus
cluster. At a radius of 10--20 kpc,  the Mg abundance of the 
Centaurus cluster is a factor of 2 larger than that of M 87 (Figure \ref{mgo}).
The stellar metallicity gradient of  M 87 is also steeper than that of 
NGC 4696, the cD galaxy of the Centaurus cluster.
The agreements of between the Mg abundances with
stellar metallicity  at the central regions indicates that in these
regions, the gas  is  dominated by gas ejected from the cD galaxies.

The Mg/O ratio of the gas of the Centaurus cluster
 is 1.2 $\pm$ 0.3 of the solar ratio on average as determined with the
 multi-temperature APEC model.
This value is close to  that of the ICM within 4$'$ of
  M 87, which is  1.3$\pm$ 0.1  solar ratio (Matsushita et al. 2003),  and  
also to the value of 1.3$\pm 0.2$ solar ratio of NGC 4636 (Xu et al. 2003) derived from the RGS.
The observed Mg/O ratios of the gas at the cluster centers
are  close to [Mg/O] of the Galactic stars (Edvardsson et al. 1993).
Therefore, at least for the Mg/O ratio, the Galactic SN II and 
those trapped in stars in the cD galaxies have no difference.

\subsection{The Fe abundance profile and SN Ia contribution}

\begin{figure}[]
\centerline{
\resizebox{8cm}{!}{\includegraphics{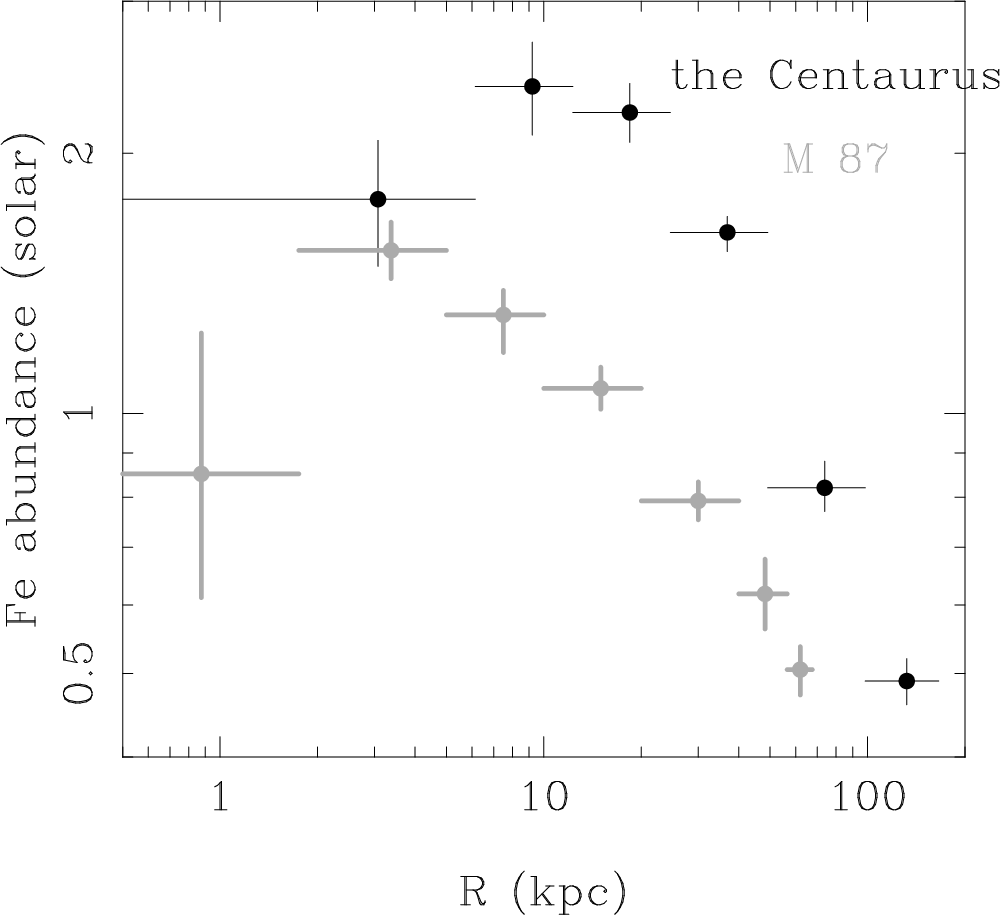}}}
\caption{The Fe abundance of the ICM of the center of Centaurus
 cluster (black) and M 87 (gray). }
\label{mgas0}
\end{figure}

The Fe in the gas at the core of clusters is also a mixture of those in the ICM and the recent
supply from the cD galaxy.
The latter contains Fe synthesized by SN Ia  and  those coming from 
stars  through stellar mass loss, since Fe is synthesized by both SN Ia
and SN II.

The Fe abundance of the ICM of the Centaurus cluster is systematically
higher than that around M 87 at a same distance from the cluster center
(Figure \ref{mgas0}).
The peak  Fe abundance of the Centaurus cluster is 2.3 solar.
This value is significantly higher than that of  M 87, which is 1.6 solar.
Assuming that the O/Fe ratio of SN II is 3,
 the central Fe abundances from SN Ia of the Centaurus cluster and M 87 
become 2 and 1.3 solar, respectively.
 Since the SN Ia contribution to the Fe abundance is dominant, the
uncertainty in the assumption for the O/Fe ratio does not change the result very much.

\begin{figure}[]
\centerline{\resizebox{8.1cm}{!}{\includegraphics{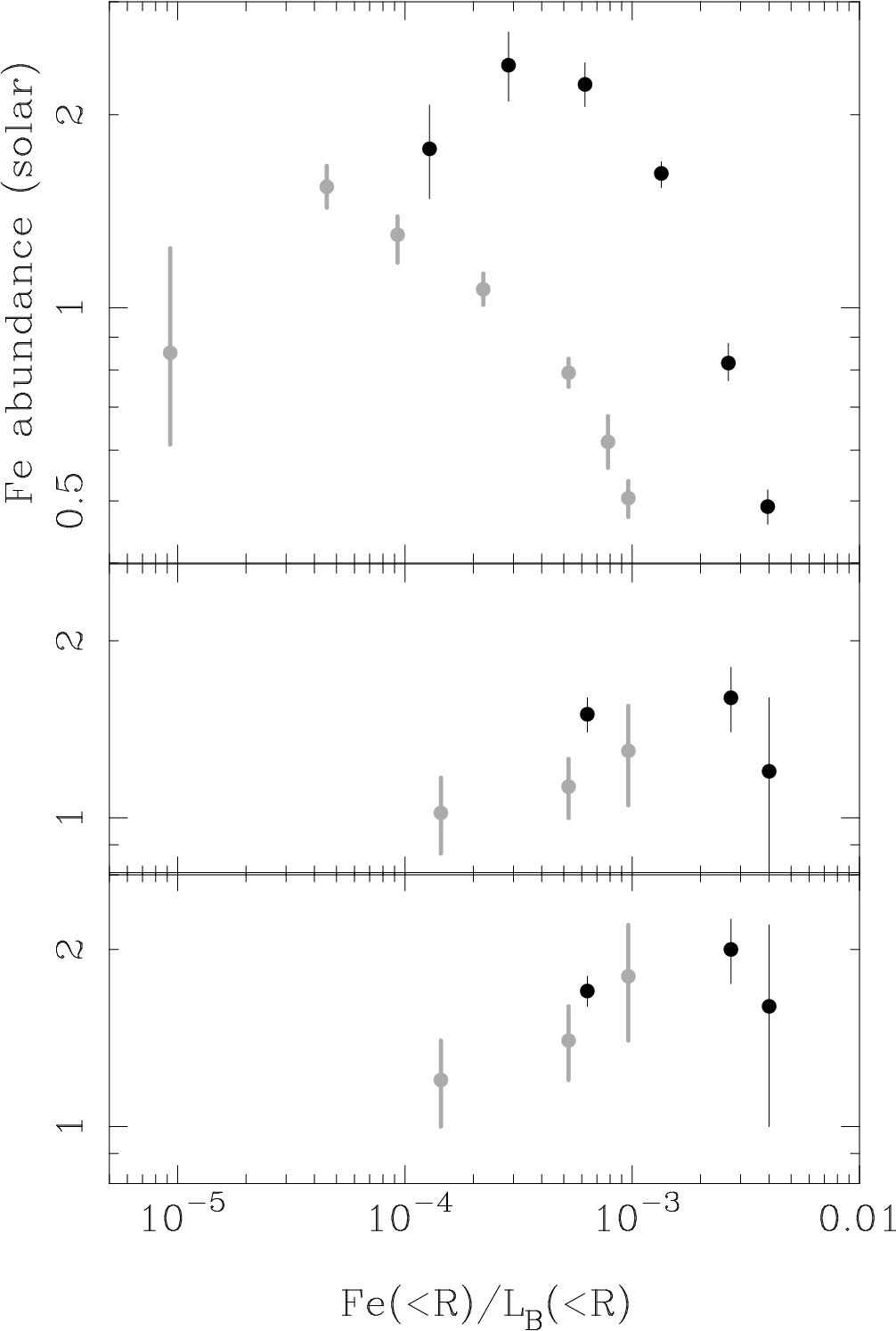}}
}  
\caption{The Fe abundance (the upper panel) and   $\rm{(Fe/Si)}_{\rm{SN Ia}}$ (the middle and the bottom panel)
in units of the solar values are
plotted against the ratio of the Fe mass to B-band luminosity  of the cD galaxy (IMLR),
for M 87 (gray circles) and the Centaurus cluster (black circles), when adopting the abundance pattern
of SN II from the average value of Galactic metal poor stars (the middle panel; Clementini et al. 1999) and from
nucleosynthesis model (the bottom panel) by Nomoto et al. (1997) assuming Salpeter's initial mass function (Iwamoto et al. 1999). }
\label{mgas}
\end{figure}

The Fe abundance of gas from SN Ia in an elliptical galaxy is 
proportional to $M^{\rm Fe}_{\rm SN} \theta_{\rm SN}/\alpha_*$(see
Matsushita et al. 2003 for details).
Here,  $M^{\rm Fe}_{\rm SN}$ is the mass of Fe synthesized by one SN Ia,
$\theta_{\rm SN}$ is SN Ia rate, and $\alpha_*$ is stellar mass loss rate.
 The higher Fe abundance of the gas in  the central region of the
Centaurus cluster indicates a higher $M^{\rm Fe}_{\rm SN} \theta_{\rm SN}/\alpha_*$.
 The cooling flow rate derived from the standard cooling flow model
is higher in the Centaurus cluster (Allen \& Fabian 1994) than that of the M 87 (Matsushita
et al. 2002).
 B\"ohringer et al. (2004)  questioned the standard cooling flow
model comparing the Fe mass profile of 4 clusters of
galaxies with cooling cores including M 87 and the Centaurus cluster.
They concluded that we need long enrichment times ($>5$ Gyr) in order to
accumulate the observed Fe abundance peak even if the SN Ia rate was
larger in the past.

As shown in Figure 18, a difference between M 87 and the Centaurus cluster is 
the ratio of the  Fe mass to the stellar B-band luminosity  of the cD
galaxy(IMLR).
The total B-band luminosity and the effective radius were calculated
using Prugniel and Heraudeau (1998), and the effect of the Galactic extinction
was corrected using Schlegel et al. (1998).
The IMLR of the Centaurus cluster is much higher than that of the M 87,
especially within the central region.
Since the contribution of the SN Ia to  Fe in the ICM is dominant,
this indicates that accumulation time scale of SN Ia is higher in the
Centaurus cluster than that in M 87, 
although we do not know the actual gas flow
rate due to the cooling flow problem.
The longer accumulation time scale in the Centaurus cluster means 
that  $M^{\rm Fe}_{\rm SN} \theta_{\rm SN}/\alpha_*$  was higher in the past.
In other words,  the Fe mass synthesized by a SN Ia was higher in the past or
the ratio of SN Ia rate to stellar mass loss rate was higher in the past 
as suggested in Renzini et al. (1993).

\subsection{Abundance pattern of SN Ia }

\begin{figure}[]
\centerline{
\resizebox{8.1cm}{!}{\includegraphics{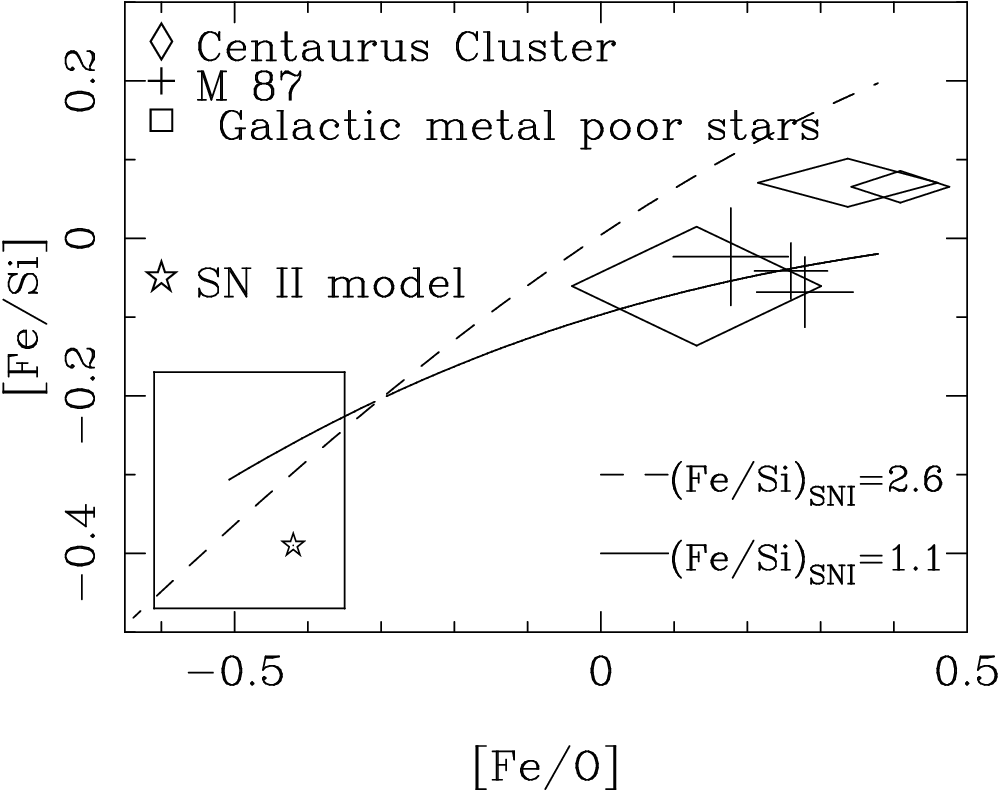}}
}
\caption{[Fe/Si] of the ICM of the Centaurus cluster (diamonds) and M 87
 (crosses; Matsushita et al. 2003) are plotted against [Fe/O].
The average value of Galactic metal poor stars (Clementini et al. 1999;
 open square)  and  the abundance ratio of SN II model using the
 nucleosynthesis model (asterisk) derived in Nomoto et al. (1997), assuming
 Salpeter's IMF (Iwamoto et al. 1999) are shown.
 The solid line and
 dashed line represent the relation of the abundance pattern synthesized by
 SN Ia with Fe/Si=1.1 (the best fit relation of M 87) and Fe/Si=2.6 (W7 ratio;
 Nomoto et al. 1984), respectively.
}
\label{osife}
\end{figure}

 Si is both synthesized by SN Ia and SN II, and the abundance
pattern of  O, Si, and Fe strongly constrains the nucleosynthesis  of the
SN Ia when Fe is dominated by SN Ia.
  Figure \ref{osife} summarizes the  ratio of these three elements in the ICM
       of the Centaurus cluster and     M 87 (Matsushita et al. 2003).
   As in M 87, the data of the Centaurus cluster cannot be explained by
       a sum of SN II abundance pattern in the Galactic metal poor stars (Clementini
       et al. 1999), and SN Ia
       abundance pattern by the W7 model.   The Si abundances are clearly higher than
       the sum of the two.

 At the cluster center, the Fe/Si ratio and the Fe/O
ratio of the gas of the Centaurus cluster are systematically higher than
those of M87, respectively.
       The main difference between M 87 and the Centaurus Cluster is the
       IMLR as described in the previous section. 
       In figure \ref{mgas},  $ \rm{(Fe/Si)}_{\rm{SN Ia}}$ is also
       plotted against IMLR.
The derived  $ \rm{(Fe/Si)}_{\rm{SN Ia}}$ value is much smaller  than the
       value of 2.6 in units of the solar ratio which is
       predicted by the W7 model.
Regions with larger IMLR have 
       larger values of  $\rm{(Fe/Si)}_{\rm{SN Ia}}$, when
   adopting the SN II pattern of either the Galactic stars
       (Clementini et al. 1999), or the
       nucleosynthesis model (Iwamoto et al. 1999), 
         A  larger  IMLR corresponds to a longer accumulation time scale.
       For  M 87, it needs only a few Gyr to accumulate the observed gas
      mass within 10kpc where the gas from the cD galaxy is dominant (Matsushita et al. 2002).
       In contrast, in the center of the Centaurus cluster,  
         ejecta of SN Ia are accumulated over a  much longer time scale than
      that of M 87.
     Therefore, the correlation between  $\rm{(Fe/Si)}_{\rm{SN Ia}}$ and IMLR
     support the suggestion  by Finoguenov et al (2002) and Matsushita
       et al. (2003) from the M 87 observation
     that  $ \rm{(Fe/Si)}_{\rm{SN Ia}}$ depends on the age of the system.
 Matsushita et al. (2003) discovered
that the Si/O ratio of the Galaxy also increases  systematically toward metal rich stars.
This result indicates that there is a source of high Si/O ratio in our
Galaxy at present and supports the scenario suggesting a  dependency of the Fe/Si ratio of
ejecta of SN Ia on the age of the system.

\subsection{The hydrogen column density and the cooling flow model}

In the ASCA era, in order to explain observed  fluxes of soft energy
photons which are much smaller than the expected value from the standard
cooling flow model,  absorption  within cluster cores was discussed
(Allen et al. 2001).
In order to obscure a cooling flow in the Centaurus cluster from the standard model,
we need at least additional column density around $2\times
10^{21}\rm{cm^{-2}}$.
However, with the XMM-Newton observations,  the observed hydrogen column
density  is too low to conceal the cool component from the cooling flow model
(Matsushita et al. 2002, B\"ohringer et al. 2002).
The observed hydrogen column density of the Centaurus cluster is also consistent
with the Galactic value, $8.06\times 10^{20}\rm{cm^{-2}}$ over the 
whole field of view.
This means that  the soft energy photons from the Centaurus cluster are not obscured.

\section{Conclusion}

With the XMM-Newton observation,  the abundance pattern of O, Mg, Si, and Fe of the ICM of the
Centaurus cluster are derived.
 At the very center, the Mg and O abundances are close to the stellar
 metallicity.
This result indicates that the gas at the very center is dominated by
the supply of the stellar mass loss from the cD galaxy.

The abundances of Si and Fe are close to each other in units of the solar
abundance, while the O and Mg abundances are smaller, like in the centers of
other clusters and groups of galaxies.
In order to explain the abundance pattern, the SN Ia ejecta from the cD
galaxy should have a higher Si/Fe ratio than the standard W7 model by Nomoto et
al. (1984).

There are small differences in the Fe abundances and the Fe/Si ratio
in the ICM between the Centaurus cluster and M 87 (Matsushita et
al. 2003). These differences and   the higher IMLR in the
Centaurus cluster may reflect a higher accumulation time scale of SN
Ia.  In order to confirm this  suspicion, we will further study
abundance patterns of the ICM of number of clusters and groups of
galaxies
with Suzaku which has a better energy resolution at the O lines.

\begin{acknowledgements}

 This work was supported by Humboldt foundation.
 The paper is based on observations obtained with XMM-Newton, an ESA science mission with instruments and contributions direct
by funded by ESA Member States and the USA (NASA).
 The XMM-Newton project is supported by the Bundesministerium f\"ur Bildung und Forschung, Deutsches Zentrum  f\"ur Luft und Raumfahrt (BMBF/DLR), the Max-Planck Society and the Haidenhain-Stiftung.

\end{acknowledgements}

\end{document}